\documentclass[times,twocolumn,final]{elsarticle}

\usepackage{medima}
\usepackage{framed,multirow}
\usepackage{tabu}
\usepackage{xcolor}
\usepackage{amsmath}
\usepackage{amssymb}
\usepackage{latexsym}
\usepackage{pifont}
\newcommand{\xmark}{\ding{55}}%

\usepackage{url}
\usepackage{xcolor}

\usepackage{hyperref}

\definecolor{newcolor}{rgb}{.8,.349,.1}

\journal{Medical Image Analysis}

\begin{document}

\verso{Given-name Surname \textit{et~al.}}

\begin{frontmatter}

\title{FiAt-Net: Detecting Fibroatheroma Plaque Cap in 3D Intravascular OCT Images}

\author[1]{Yaopeng \snm{Peng}}

\author[2]{Zhi  \snm{Chen}
}

\author[2]{Andreas  \snm{Wahle}}
\author[3]{Tomas  \snm{Kovarnik}}

\author[2]{Milan \snm{Sonka}}

\author[1]{Danny Z. \snm{Chen}\corref{cor1}}\cortext[cor1]{Corresponding author: D. Chen (dchen@nd.edu) }

\address[1]{Department of Computer Science and Engineering, University of Notre Dame, Notre Dame, IN 46556, USA}
\address[2]{Department of Electrical and Computer Engineering, University of Iowa, Iowa City, IA 52242, USA}
\address[3]{Second Department of Medicine, Department of Cardiovascular Medicine, First Faculty of Medicine, Charles University in Prague and General University Hospital in Prague, Prague, Czech Republic}

\received{1 May 2013}
\finalform{10 May 2013}
\accepted{13 May 2013}
\availableonline{15 May 2013}
\communicated{S. Sarkar}

\begin{abstract}
The key manifestation of coronary artery disease (CAD) is development of fibroatheromatous plaque, the cap of which may rupture and subsequently lead to coronary artery blocking and heart attack. As such, quantitative analysis of coronary plaque, its plaque cap, and consequently the cap's likelihood to rupture are of critical importance when assessing a risk of cardiovascular events. This paper reports a new deep learning based approach, called FiAt-Net, for detecting angular extent of fibroatheroma (FA) and segmenting its cap in 3D intravascular optical coherence tomography (IVOCT) images. IVOCT 2D image frames are first associated with distinct clusters and  data from each cluster are used for model training. As plaque is typically focal and thus unevenly distributed, a binary partitioning method is employed to identify FA plaque areas to focus on to mitigate the data imbalance issue. Additional image representations (called auxiliary images) are generated to capture IVOCT intensity changes to help distinguish FA and non-FA areas on the coronary wall. Information in varying scales is derived from the original IVOCT and auxiliary images, and a multi-head self-attention mechanism is employed to fuse such information. Our FiAt-Net achieved high performance on a 3D IVOCT coronary image dataset, demonstrating its effectiveness in accurately detecting FA cap in IVOCT images.
\end{abstract}

\end{frontmatter}


\section{Introduction}
\label{sec1}

Cardiovascular disease is the leading cause of death in the US, accounting for 20\% of all deaths nationwide~\citep{national2023multiple, tsao2022heart}. Coronary artery disease (CAD) is the most common among cardiovascular diseases, CAD develops when arteries supplying blood to the heart muscle become narrowed and  plaque regions develop affecting coronary wall health~\citep{malakar2019review}. The most frequent major cardiac events are due to plaque rupture. A commonly-used clinical treatment is performing baloon angioplasty, frequently followed by  placing a stent to help prevent the artery from re-closing. Thus, detecting vulnerable plaque regions prior to their rupture is critical to administer early treatment.  Fibroatheroma (FA) is a major precursor lesion to acute coronary syndromes~\citep{kolodgie2001thin}. FA is enclosed by a  fibrous cap covering a lipid-rich core containing inflammatory cells and necrotic debris. Identification of FA cap can help cardiologists decide whether further treatment, such as percutaneous coronary intervention (PCI), is necessary.

\begin{figure}[t]
\centerline{\includegraphics[width=0.7\columnwidth]{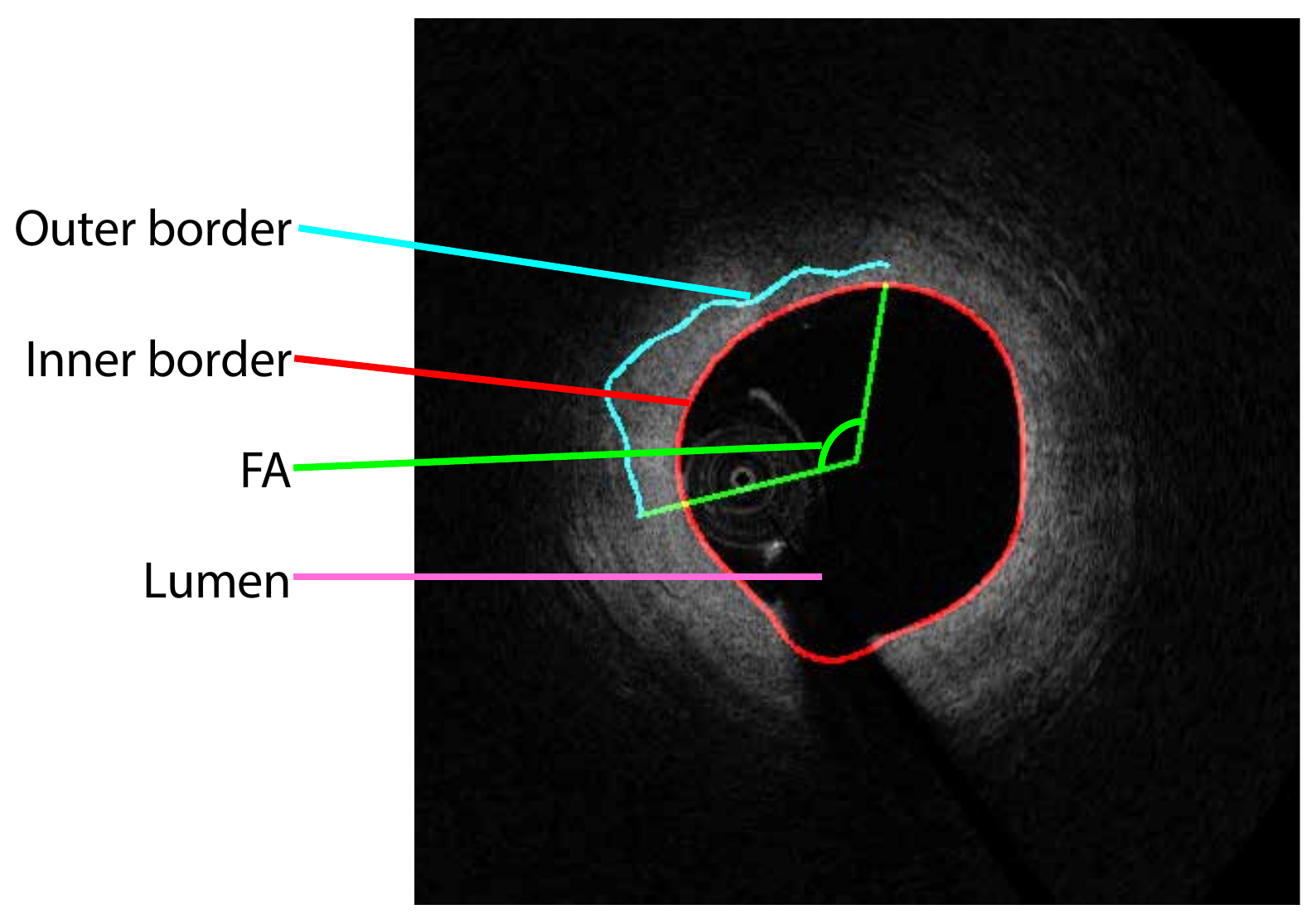}}
\caption{Illustrating  the angular coverage of a fibro-atheromatous plaque cap (FA-cap angle) in a 2D IVOCT frame. The red curve marks the inner border between the lumen and vessel wall; the blue curve marks part of the outer border between the vessel wall and periadventitial tissues; the green angular range shows the FA radial segment.}
\label{tcfa}
\end{figure}

In recent years, intravascular optical coherence tomography (IVOCT) has been used increasingly to assist the detection of FA, segmenting its cap, and guide PCI. IVOCT is a minimally-invasive imaging modality that enables tissue visualization {\it in vivo} at near-histology resolution. Compared to coronary angiography and intravascular ultrasound (IVUS), IVOCT provides more detailed information and higher resolution of the vessel walls. During IVOCT imaging, a thin catheter is inserted into the artery and pulled back while capturing images along its axis in an acquisition speed of 100-160 frames per second and a pullback speed of 15-25 mm per second. By emitting and receiving near-infrared light at each angular direction, an array of axial lines (A-lines) can be obtained. Multiple A-lines are combined to create a 2D image frame (B-scan) of the tissue. A series of adjacent B-scans is generated to form a 3D image stack. A stack of adjacent cross-sectional frames along the length of the assessed artery segment is reconstructed by converting the intensities and stacks of all A-lines into a grayscale image representation~\citep{Zahnd} (e.g., see Fig.~\ref{tcfa}).

On a healthy artery, a triple-layer structure consists of the intima, media, and adventitia. 
The lumen and intima are separated by the inner border, while the adventitial and periadventitial layers are separated by the outer border. The tissue structure between the inner and outer borders forms the region of interest (RoI) (e.g., see Fig.~\ref{tcfa}), which may encompass the lipid core of atherosclerotic plaque. The stability of the fibrous cap is important in determining the risk of plaque rupture, which can cause forming of blood clots and ensuing blockage of blood vessels.

The common practice of detecting FA often first segments the layer structure of the artery walls and then extracts its angle-specific features. In this study, we formulate FA detection as an angle prediction problem (e.g., see Fig.~\ref{tcfa}) instead of a voxel prediction problem, for the following reasons. (i) The aim of detecting FA areas is to provide physicians with recommendations on whether early intervention, usually cardiovascular stent implantation, is needed for vessel treatment. Thus, angular-level information, rather than voxel-level information, is more critical for decision-making. (ii) Annotating an angular area for FA is of a much lower cost and easier compared to voxel-wise annotation. Moreover, there are no obvious intensity regions and boundaries to delineate FA compared to other types of lesion annotations in medical imaging. (iii) The characteristics of FA are such that the brightness and shadows of the fibrous cap are different compared to non-FA regions (e.g., radial-axial-wise rather than voxel-wise). 

Although IVOCT is capable of capturing substantial anatomical details of vessel walls, learning-based FA detection approaches commonly suffer from the sparse distribution of FA. During IVOCT pullback, FA presence usually accounts for only a small portion of the entire pullback, i.e., among the frames containing FA, the target angles in a frame may cover just a small range. The scarcity of FA makes the data distribution extremely imbalanced, giving a high chance of mis-detection of the (small) FA areas. Models trained on such data are likely to incur a high false negative rate and may miss a large proportion of areas containing FA.

Besides the aforementioned challenge, the borders of different artery layers in IVOCT images and the boundaries between the vessel walls and background are often very unclear. Further, numerous artifacts may occur during the IVOCT imaging process. All these challenges pose difficulties to IVOCT image analysis and FA detection.  Table~\ref{vanilla_seg} provides the results of various known deep learning (DL) models for detecting FA angles on our IVOCT dataset, which show that the F1-scores of even the very recent DL segmentation methods are not satisfactory.

\setlength{\tabcolsep}{1pt}
\begin{table}[t]
\centering
\caption{Results of various known DL segmentation methods for detecting FA angles as applied to the analyzed coronary dataset.}
\label{vanilla_seg}
\begin{tabular}{r|l|l|l}
Method &  F1 & AUC  & Accuracy  \\
\hline		
U-Net~\citep{ronneberger2015u} & 63.45  & 79.81 & 80.72 \\			
TransUNet~\citep{chen2102transformers} & 63.82  & 83.07 & 81.32\\		
PraNet\citep{fan2020pranet} & 65.19  & 81.37 & 81.29\\		
Attention U-Net~\citep{oktay2018attention} & 66.97  & 85.77 & 82.05 \\
Swin-Unet~\citep{cao2023swin} & \textbf{69.95} & \textbf{88.80} & \textbf{83.45}

\end{tabular}
      
\end{table} 

In this paper, we propose a new DL approach, called FiAt-Net, for effectively identifying angle areas containing FA in 3D IVOCT images (i.e., the FA-cap angles, if any, in each 2D frame of a 3D image, as illustrated in Fig.~\ref{tcfa}). The main steps of our pipeline are as follows. (1) Pre-process the input 3D IVOCT image using a dynamic programming algorithm to detect the luminal and abluminal borders, removing irrelevant background and noise areas. An ablation study in Section~\ref{li_process} demonstrates the effectiveness of the pre-processing in boosting the performance. (2) Cluster the 2D frames of all 3D training images using an auto-encoder mechanism and organize them into multiple clusters, so that frames similar to one another are grouped together and dissimilar ones are separated; this enables different types of plaques to be sampled uniformly, thus mitigating the imbalanced distribution issue and improving performance. (3) Build a binary tree to help narrow down the search for FA areas and focus the attention on the target areas of interest. Consequently, this method amplifies the model's 
focus on FA regions while attenuating the influence of the predominant non-FA regions. (4) Explicitly generate additional image representations (called auxiliary images) to capture various intensity changes along the radial directions of vessel walls and the brightness and shadows between the lumen and abluminal surfaces, for distinguishing FA and non-FA areas. This provides more informative clues for the model to discriminate FA and non-FA. Moreover, we incorporate features in varying scales from the original image and auxiliary images, and employ a multi-head self-attention mechanism to fuse these features. 

\section{Related Work}
Many automated methods have been proposed to detect FA in IVOCT. These methods fall into two main categories. (A) Image-level detection: Such methods classify image frames as containing or not containing FA. Min et al.~\citep{Min} proposed a DL model to classify frames as with or without OCT-captured FA. Jun et al.~\citep{Jun} gave methods to classify FA using various machine learning classifiers (e.g., feed-forward neural network (FNN), K-nearest neighbors (KNN), random forest (RF), and convolutional neural network (CNN)), and identified a classifier with the best classification accuracy. In \cite{gessert2018automatic}, a two-path architecture was proposed to simultaneously utilize the polar and Cartesian representations of each frame, which were concatenated for binary classification. (B) A-line based detection: These methods classify, localize, and segment FA based on A-lines, which are individual beams used to create OCT images, neglecting the spatial context. Shi et al.~\citep{shi2018vulnerable} and Kolluru et al.~\citep{Kolluru} classified each A-line of a frame and predicted the extent of FA lesions. Liu et al.~\citep{Liu} detected lesion locations by classifying region proposals generated by a DL network. Liu et al.~\citep{Liu_S} proposed a single unified salient-regions-based CNN to recognize vulnerable plaques, utilizing multi-annotation information and combining prior knowledge of cardiologists. Li et al.~\citep{Li} developed a method to segment vulnerable cardiovascular plaques by constructing a Deep Residual U-Net with a loss function consisting of weighted cross-entropy loss and dice coefficient. Lee et al.~\citep{Lee} used the DeepLab-v3 plus model~\citep{chen2018encoder} to classify lipidous plaque pixels, detected the outer border of the fibrous cap using a special dynamic programming algorithm, and assessed cap thickness. Abdolmanafi et al.~\citep{Abdolmanafi} trained a random forest using CNN features to distinguish normal and diseased arterial wall structures; the tissue layers in normal cases and pathological tissues in diseased cases were extracted by a fully convolutional network (FCN) to classify the lesion types.

Although the above methods are effective in determining whether a frame contains FA and localizing the areas of FA, they still suffer from the fact that the sparse occurrences of FA can hinder DL models in detecting FA on clinical datasets where most subjects or frames do not contain FA lesions. To address this challenge, we propose a new approach that can detect FA areas even when FA is sparsely distributed among the frames of 3D IVOCT images. 

\section{Method}
In this section, we present 
our FiAt-Net approach, which contains three main stages. (1) {\it Apply a series of pre-processing steps to clean the input frames by removing some noise and background areas.} (2) {\it Based on their latent feature vectors, divide all the frames  into different clusters so that similar frames fall within the same cluster, and build training batches by randomly sampling from each cluster.} Since there can be multiple types of disease regions in IVOCT images (e.g., calcification), building training batches in this way enables the model not to lean toward a specific type of region and degenerate. 
(3) {\it Construct a hierarchical structure that gradually narrows down the FA areas using a binary partition tree, and utilize a multi-head attention mechanism to incorporate features extracted at different tree levels.} The binary partition tree method enables the network to focus on target FA areas, feeding the model with more target areas and less noise and background areas. Moreover, it reduces the negative effect of sparse FA distribution. 
The multi-head attention mechanism incorporates features of different scales from the raw image and from additional image representations (called auxiliary images) that we build to capture intensity changes for distinguishing FA and non-FA areas.

\subsection{Pre-processing}
\begin{figure}
\centerline{\includegraphics[width=0.85\columnwidth]{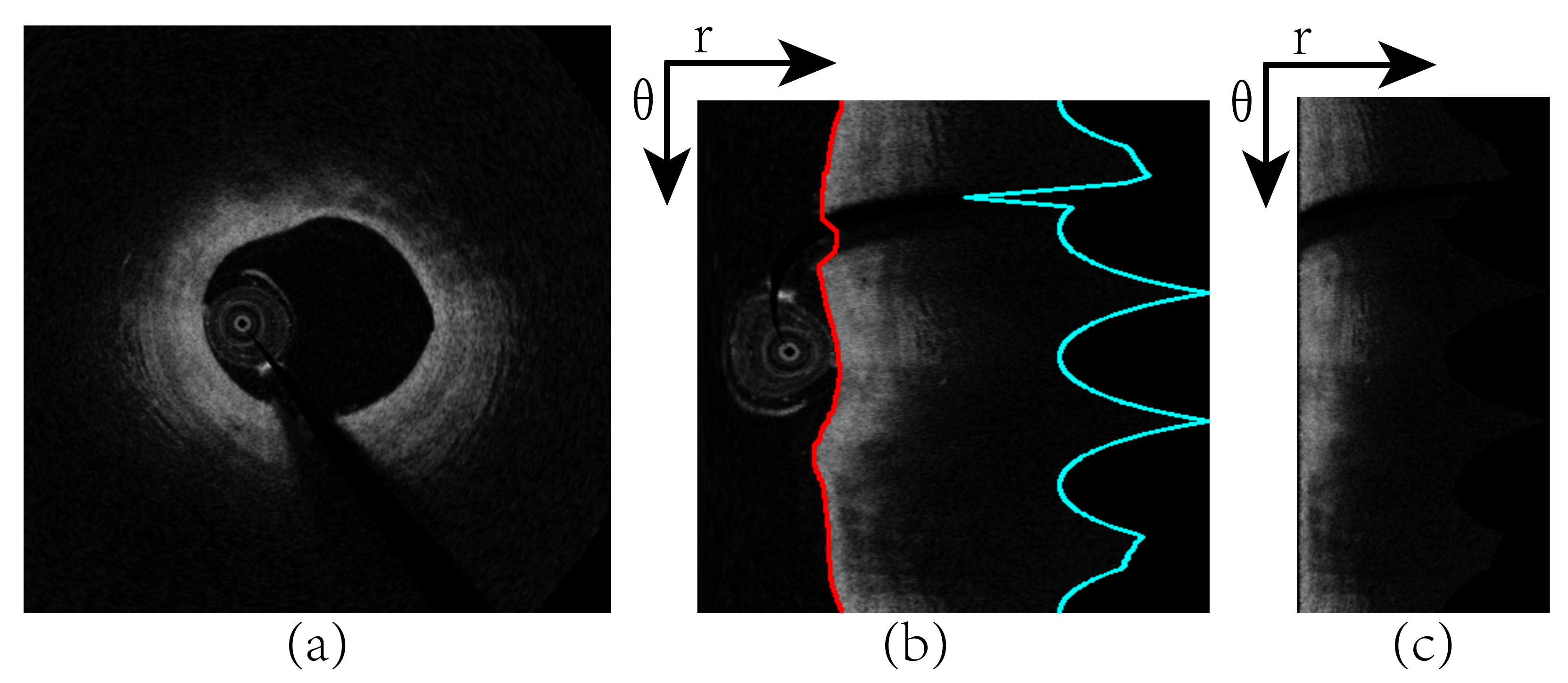}}
\caption{Illustrating our pre-processing. (a) An original OCT frame in the Cartesian domain; (b) the detected luminal (red line) surface; (c) the resulted frame of the pre-processing. We first employ the pre-processing step in~\citep{zahnd2017contour} to detect the lumen border (the red curve in 
panel (b)). The boundary between the RoI area and background area (all zero) is detected following the Cartesian-Polar conversion. Each row is subsequently shifted so that the luminal border forms a straight vertical line (the left boundary) in the polar-coordinate frame (panel (c)). This step removes catheter artifacts and blood remnants in the lumen area. Next, we use the row that has the longest lumen-background distance (red-to-blue) as the base and right-pad (with 0's) the other rows whose lumen-background distances are shorter than the longest distance. Finally, we resize each image to the size of $360\times 128$, ensuring that each image has the same size and the original pixel density is not affected.}
\label{preprocess}
\end{figure}

In IVOCT images, FA is manifested by a cap, a lipid core, and an increase of inflammatory cells in the arterial wall. To retain only the areas of interest, we first remove noise and lumen areas in IVOCT images before locating FA.

Given an input frame in the Cartesian domain, we first convert the frame from the Cartesian domain to the polar domain by sampling along each of 360 angular directions, anchoring at the frame center, starting at degree $0$ (3 o'clock) and proceeding clockwise (see Fig.~\ref{preprocess}(b), where the blue curve is the outer boundary of the polar image). Next, we apply the first-order $x$-derivative to enhance the polar frame, as:
\begin{eqnarray}
	I' = I \circledast {\bf k},
\end{eqnarray}
where $I$ and ${\bf k}$ denote the polar frame and first-order $x$-derivative Gaussian kernel with standard deviation $\sigma$, respectively, and $\circledast$ denotes convolution. Next, to detect the lumen-intima border and remove some background and noise areas, we apply the dynamic programming method in \citep{wang2012volumetric}. Finally, we extract the RoI (the region between the red and blue curves in Fig.~\ref{preprocess}(b), i.e., the pixels excluding the lumen) by shifting each row so that the luminal border corresponds to a straight vertical line (the left boundary) in the polar frame (e.g., see Fig.~\ref{preprocess}(c)). After this process, some noise and unrelated areas including the guide-wire, probe, and blood remnants are removed. Each frame thus pre-processed is then fed to the model for FA detection. Fig.~\ref{preprocess} illustrates the pre-processing. Note that the Cartesian-to-polar space conversion is performed only during the pre-processing steps, while the other 
operations are conducted on the polar image.

\subsection{Frame Clustering}

\begin{figure*}[h!]	
	\centering
	\centerline{\includegraphics[width=320.0pt]{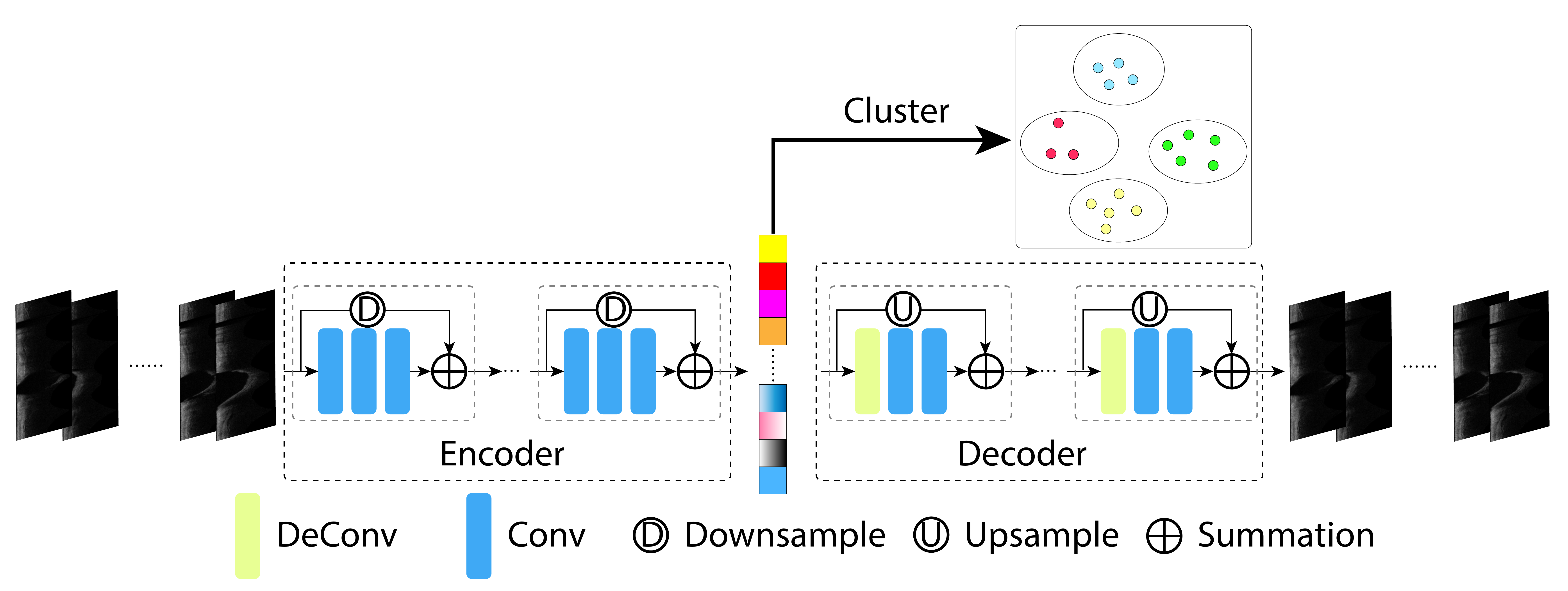}}
	
	\caption{Illustrating the frame clustering process. An auto-encoder is used to extract latent features of each frame, and these latent features are used for frame clustering.}
	\label{ImageCluster}
\end{figure*}

After the pre-processing, a simple approach is to feed all the training frames and annotations to a neural network for binary classification. However, the FA distribution in these frames may vary largely. For example, FA regions commonly account for only a small portion, and other plaque-type regions may also be present (e.g., atheroma and calcified nodules). Since we have annotations only for FA, other types of regions are all treated as negative. Furthermore, the distributions of different types of regions are unclear, which could lead the model to lean towards a non-FA type of region and degrade its performance. Instead of randomly sampling frames from all the training frames to form a training batch, stratified sampling can make the model's predictions closer to the real data distribution. For this, we apply image clustering methods to group similar frames into the same cluster. We expect that frames in the same cluster are similar, and let each training batch contain samples from each cluster.

We first apply an auto-encoder to reconstruct the polar frames, which learns to store relevant cues of each frame and discard unrelated information. The auto-encoder has two parts: an encoder that produces a compression $x$ for an input image $I$, and a decoder that reconstructs the original image taking $x$ as input. It is optimized by minimizing the sum of the reconstruction error that measures the difference between the input image $I$ and the reconstructed image, and a regularization term that mitigates overfitting. This is formulated as:
\begin{equation}
	\phi^*, \Phi^* = \arg\min_{\phi, \Phi} ({\cal {L}}(I, \hat x)+ \lambda \times \sum_{i=1}^M w_i^2),
\end{equation}
where ${\cal {L}}$ is the reconstruction loss between the input $I$ and the reconstruction $\hat x$, $\lambda$ is a scaling parameter for the regularization term that adjusts the trade-off between sensitivity to the input and overfitting, $w_i$ denotes the $i$-th parameter of the auto-encoder, and $\phi$ and $\Phi$ denote the parameters of the encoder and decoder, respectively.

To facilitate fast training and convergence of the auto-encoder, we utilize a pre-trained ResNet-101 \citep{resnet} based on ImageNet \citep{imagenet} as the encoder backbone. A lightweight decoder (ResNet-50 \citep{resnet}) is added to map the latent vectors back to the original input space.

After training the auto-encoder, its encoder part is extracted and utilized to generate a latent feature vector for each training frame. As it is difficult to pre-determine the number of frame clusters, we apply the agglomerative clustering algorithm \citep{mullner2011modern} for the grouping process. For a set of $n$ frame feature vectors, each vector is initially treated as a single cluster. We iteratively merge two clusters that are most similar to form a new, larger cluster. This process continues until all the vectors are merged into a single large cluster, resulting in a dendrogram. Finally, we select a threshold to cut the dendrogram and obtain individual clusters. For simplicity, each cluster is represented by its centroid, and the distance between two clusters is measured by the cosine distance between their centroids.

After the frame clustering of the training images, we evenly sample frames randomly from each cluster to build every training batch, thus making the samples more diverse. Furthermore, this training scheme is capable of preventing the model from leaning towards any type of non-region and degrading its performance.  Fig.~\ref{ImageCluster} shows the frame clustering process.

\subsection{Binary Partition and FiAt-Net Model}

\begin{figure}[t]
	\centerline{\includegraphics[width=0.6\columnwidth]{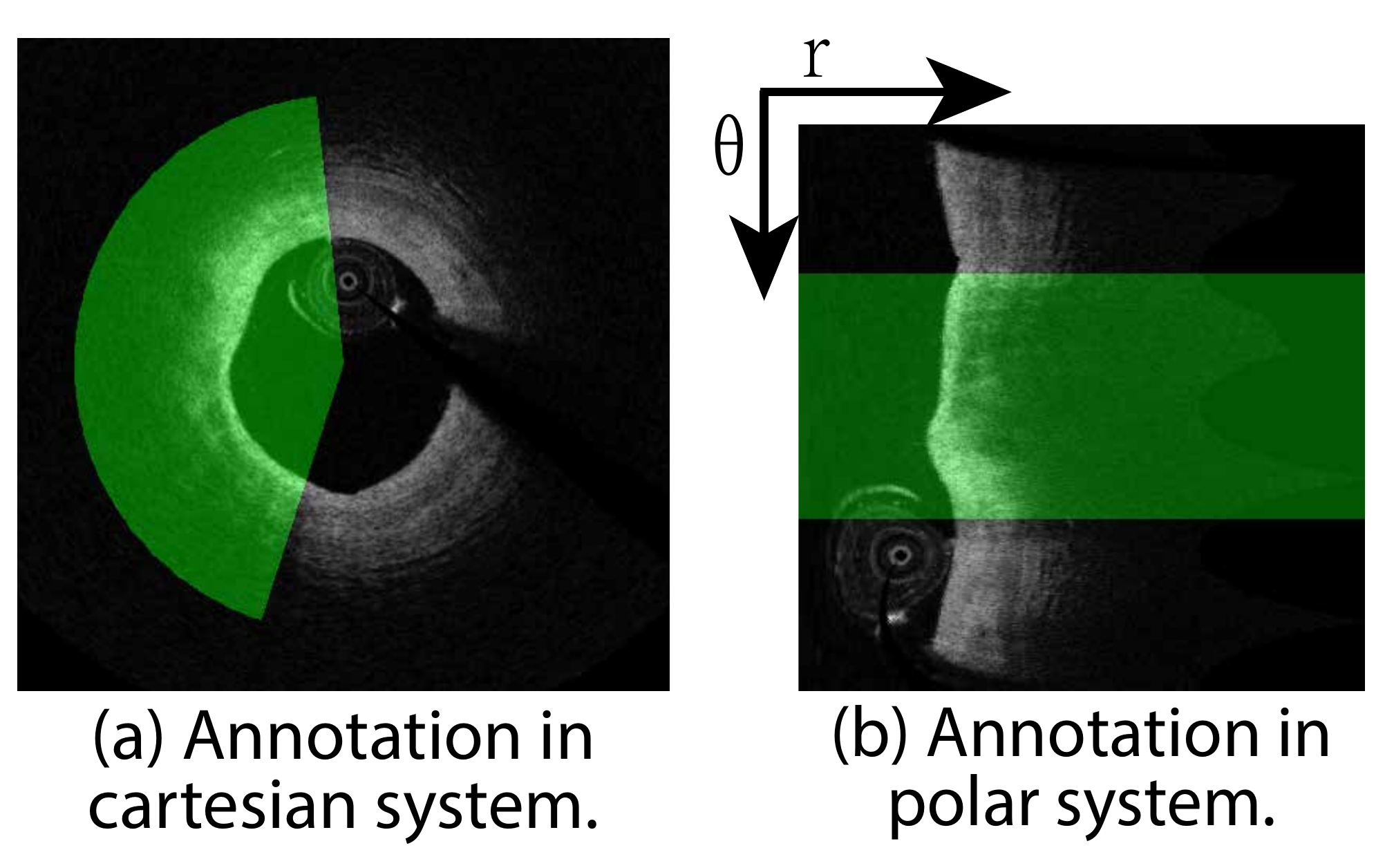}}
	\caption{Illustrating FA annotation in (a) the Cartesian domain and (b) the polar domain. The green areas represent FA ranges.}
	\label{Annotation}
\end{figure}

In IVOCT images, FA refers to areas whose brightness, shadows, and the shape of the border behind the caps are different from non-FA regions of the vessels. Experts typically mark an FA area in a frame as an angular range in the Cartesian domain or a vertical range in the polar domain (e.g., see Fig.~\ref{Annotation}). The vulnerable frames with FA often account for only a small portion of all frames. Even among the vulnerable frames, the FA target constitutes only a small fraction. This sparse distribution of FA poses a challenge for the model to learn effective features.

\begin{figure*}[h!]
\centerline{\includegraphics[width=1.6\columnwidth]{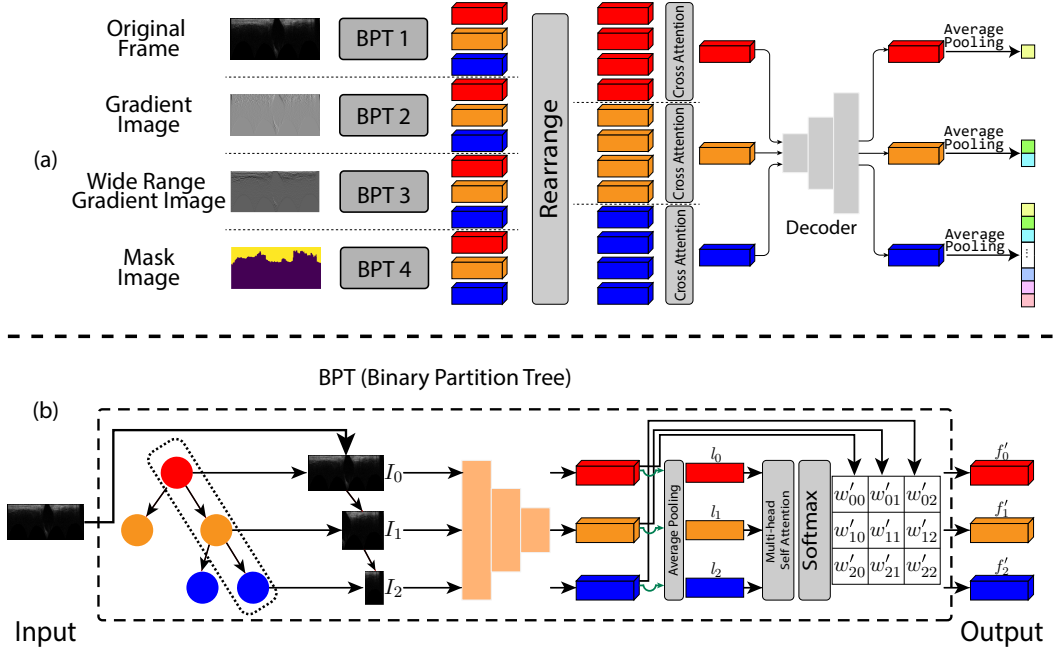}}
\caption{Illustrating the FiAt-Net. (a) The overall process. For an input frame and each of its three auxiliary images, we extract features at different levels (in this example, marked by \textcolor{red}{red}, \textcolor{orange}{orange}, and \textcolor{blue}{blue}, respectively) of its BPT (binary partition tree). Next, we aggregate features of the same level from all the four BPTs. Finally, the aggregated features of each level are used to generate the output and compute the loss. (b) The process on one BPT. For one input frame or an auxiliary image, a root-to-leaf path is randomly selected; let its sequence of frame and sub-regions be, e.g., $S=(I_0, I_1, I_2)$. Our model takes $S$ as input, uses a multi-head self-attention mechanism to integrate their feature maps $f_0, f_1$, and $f_2$ at different levels, and outputs refined feature maps $f'_0, f'_1$, and $f'_2$. We only illustrate a path of three levels for an original frame for simplicity.}
\label{BinaryTree}
\end{figure*}

An intuitive way to address the issue of sparse FA distribution is to first differentiate the FA and non-FA frames and then focus on detecting the FA ones. Along this idea, we take a step further to continually divide each FA frame into two equal-size parts and identify FA in each part, if any.  Hence, we propose a binary partition method to gradually narrow down the FA areas. That is, we repeatedly partition a polar domain frame into two sub-regions to search for FA, as shown in Fig.~\ref{BinaryTree}(b).

We train our model to determine whether a frame $F$ contains FA. If FA is not detected, then we move on to the next frame. Otherwise, we divide $F$ equally into two sub-regions along a horizontal line. Likewise, we continually partition any of the sub-regions if FA is detected in it. By recursively performing this process, we can narrow down the search of FA step by step, until the search range is small enough.

In this FA search process, one issue is how to identify FA areas, either when a sub-region is too small (we set this as the region width less than 4 pixels) and not partitioned any further or when it does not contain FA. We use the idea of traversing a graph (more precisely, a tree), and visit the binary partition tree $T$ using a depth-first search (DFS) traversal. 
Each frame or sub-region is treated as a node $v$ in $T$, and its two partitioned sub-regions (if any) are taken as its two children. If FA is detected in a node $v$ and the region is large enough, we divide it into two equal-size parts and recursively search each of its two children. If FA is not detected, we stop the search at the current node.

We define a ``negative" threshold $\alpha$ for the search process (empirically, we set $\alpha$ = 4). When the non-FA areas in a sub-region $R$ are larger than $L-\alpha$, where $L$ is the angle range size of $R$, we consider $R$ as negative and stop the search at $R$. In the inference stage, if the confidence of non-FA in $R$ is larger than $1-\frac{\alpha}{L}$, we consider $R$ as negative and stop the search at $R$. Otherwise, we continue the binary partition at $R$ (if $R$ is large enough) and search $R$'s children. With this process, we significantly reduce the ratio of negative samples. The search process is shown in Fig.~\ref{BinaryTree}(b). A feature map is generated for the region/sub-region at each node of $T$, with higher level maps focusing more on global features and lower-level maps focusing more on local features.

Our FiAt-Net model consists of four main components (see Fig.~\ref{BinaryTree}(a)), for BPT 1, BPT 2, BPT 3, and BPT 4, which process an original frame (OF) and three types of auxiliary images of that original frame (Gradient Image (GI), Wide-range Gradient Image (WGI), and Binary Mask Image (BMI), to be discussed in Sect.~\ref{auxilary_images}), respectively. Each component takes one of inputs OF, GI, LGI, and BMI, as a BPT (binary partition tree, see Fig.~\ref{BinaryTree}(b)), and outputs features at different levels of the BPT (marked with \textcolor{red}{red}, \textcolor{orange}{orange}, and \textcolor{blue}{blue} in the example of Fig.~\ref{BinaryTree}(b), respectively). The four components have the same structure and process but do not share parameters, because the original frame and auxiliary images have different appearances and applying the four separate components yields better performance. The integration of the features attained by the four components will be discussed in Sect.~\ref{auxilary_images}.

The processes of these components are the same, as follows. For a 2D frame (an original frame or an auxiliary image), we first use a DFS-like algorithm to generate the collection $P$ of all root-to-leaf paths in its BPT $T$.  We randomly sample one path $p$ from the collection $P$ and feed information of the path $p$ to a shared encoder in each iteration of the training process.

Suppose the path $p= (v_0, v_1, \ldots, v_{m-1})$. Let the frame at the root $v_0$ of $T$ be $I_0\in R^{1\times H\times W_0}$ (at level $0$), and the sub-regions at nodes $v_1, \ldots, v_{m-1}$ be $I_1, \ldots, I_{m-1}$, respectively, with $I_i\in \displaystyle{R^{1\times H\times \frac{W_0}{2^i}}}$.

We input the sequence $S = (I_0, I_1, \ldots, I_{m-1})$ into the shared encoder, and generate the corresponding feature maps $f_0, f_1, \ldots, f_{m-1}$, 
where $f_i\in \mathbb{R}^{C\times H\times W_i}$ is for $I_i$,  $C, H,$ and $W_i$ are for the channels, height, and width of the $i$-th level feature map $f_i$ respectively, and $W_i=\frac{W_0}{2^i}$. 
We then apply a multi-head self-attention mechanism~\citep{vaswani2017attention} to integrate the features in $f_0, f_1, \ldots, f_{m-1}$ from different levels of $T$, as follows (see Fig.~\ref{BinaryTree}(b)). 

First, we perform an average pooling on each feature map $f_i$ and generate a feature vector $l_i$, with $l_i\in \mathbb{R}^C$. Then, a query $q_i$ and a key $k_i$ are derived by multiplying $l_i$ with two learnable matrices $Q$ and $K$, with $Q, K\in R^{C\times C}$, as:
\begin{eqnarray}
  q_i = l_i * Q,  \
  k_i = l_i * K. 
\end{eqnarray}
    
The relationship strength between $f_i$ and $f_j$ is computed as:
\begin{equation}
    w_{ij}=q_i * k_j^T.
\end{equation}

Afterwards, we use a softmax function to normalize the relationship strength, as:

\begin{equation}
    w'_{ij} = \frac{e^{w_{ij}}}{\displaystyle \sum_{h=0}^{m-1}e^{w_{ih}}}.
\end{equation}

The aggregated feature map $f'_i$ is then computed as:
\begin{equation}
    f'_i=\sum_{j=0}^{m-1}w'_{ij}*h(f_j),
    \label{aggregated-feature-map}
\end{equation}
where $h(f_j)$ represents the bilinear interpolation of $f_j$ to match the resolution of $f_i$.

\begin{figure}[t]
\centerline{\includegraphics[width=0.56\columnwidth]{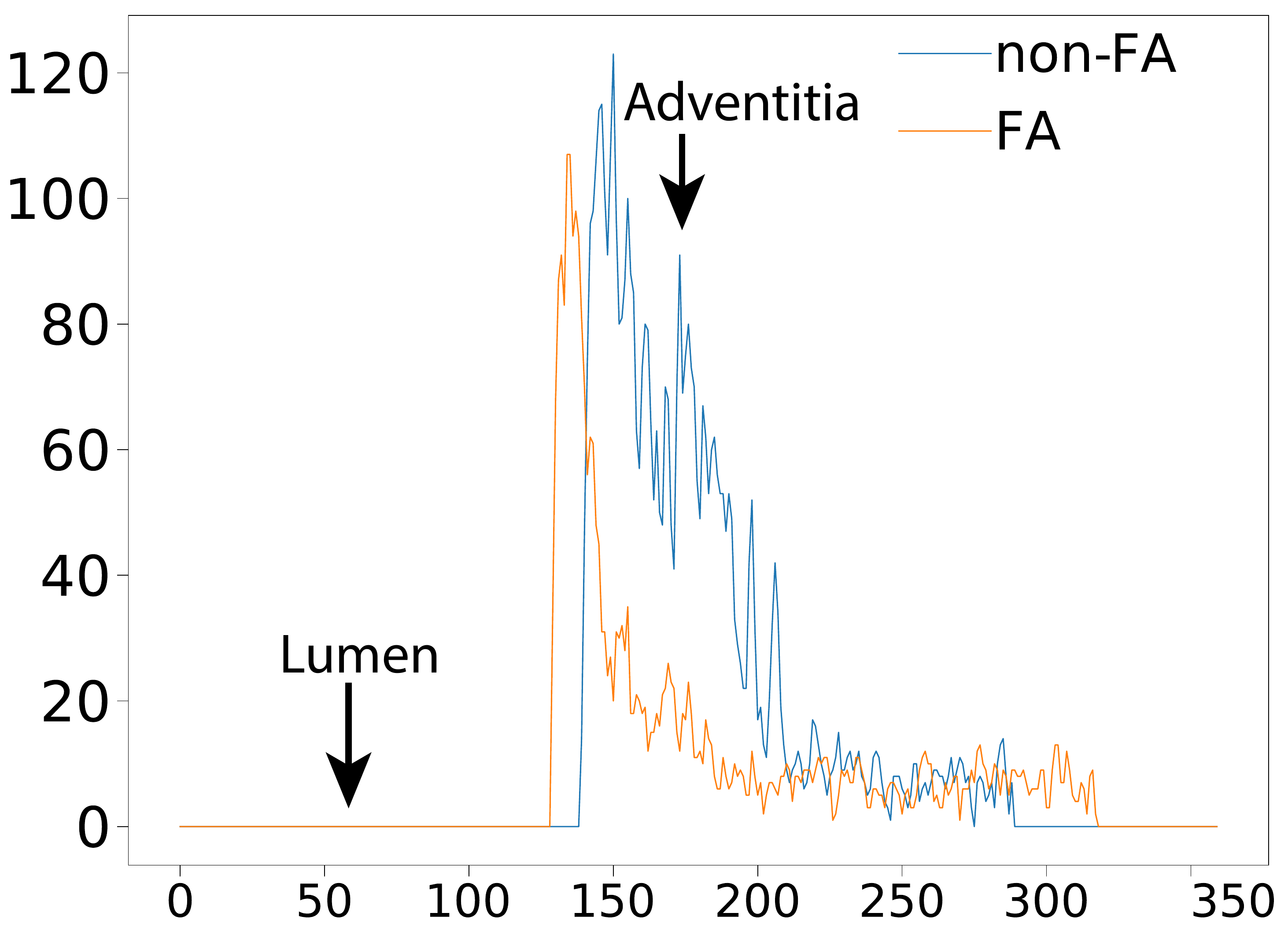}}
\caption{Illustrating intensity profiles along the radical axis in the polar domain: Intensity drops more drastically in FA areas than non-FA areas.}
\label{intensity3D}
\end{figure}

\begin{figure}[t]
	\centerline{\includegraphics[width=0.56\columnwidth]{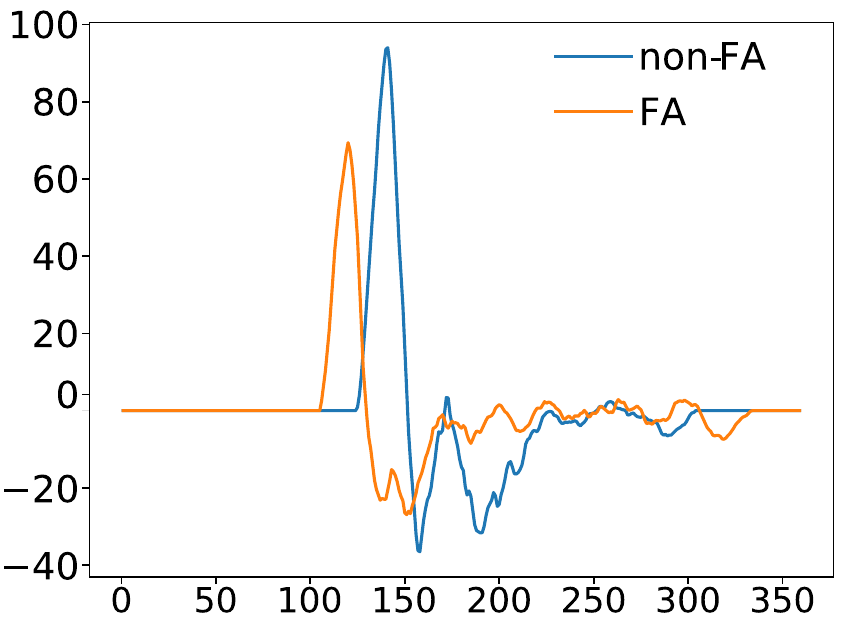}}
	\caption{Intensity change comparison in a wide range between FA and non-FA areas (for Wide-range Gradient Image (WGI)). Here, we set the range parameter $m=9$.}
	\label{diff}
\end{figure}

\subsection{Auxiliary Images and Feature Aggregation}
\label{auxilary_images}
For an original frame (OF) $I$, we construct three auxiliary images for $I$: the Gradient Image (GI), Wide-range Gradient Image (LGI), and Binary Mask Image (BMI).

\noindent
\textbf{Gradient Image (GI):} In an unfolded IVOCT frame, one important factor for determining the presence of FA is the layered structure. In healthy vessels, a clear three-layer structure is usually observable, consisting of the intima, media, and adventitia layers. However, in frames or regions with FA, this layered structure becomes less apparent. Information that helps distinguish this key difference can assist FA detection. Based on this observation, we incorporate gradient information along the radial axis. In healthy regions, the layered structure (intima, media, and adventitia) gives an intensity pattern along the radial axis in the polar domain as: dark $\rightarrow$ light $\rightarrow$ dark $\rightarrow$ light $\rightarrow$ dark. However, a FA area often exhibits more rapid intensity changes, as shown in Fig.~\ref{intensity3D}. 

Given an original frame ($OF$) of size $H\times W$, the gradient image ($GI$) is computed as:
\begin{equation}
    GI = N({\bf k}\circledast OF),
\end{equation}
where ${\bf k}$ is the convolution kernel $[-1, 0, 1]$ with a stride of $1$, $\circledast$ denotes 2D convolution, and $N(x)$ represents normalizing $x$ to the range of $(0, 1)$.

\noindent
\textbf{Long-range Gradient Image (LGI):} While the gradient is capable to capture the intensity difference between FA and non-FA, it primarily measures the local difference between two consecutive columns (e.g., see Fig.~\ref{Annotation}(b)). We need to further capture intensity changes over a wider range. Thus, we define a measure $M_I$ for the intensity difference between the left and right sides of each pixel within a specified range:
\begin{equation}
	\begin{split}
	M_I(i, j_\beta) = \overline I(i, j_\beta -& m < j \leq j_\beta) - \\ &\overline I(i, j_\beta < j \leq j_\beta +m),
	\end{split}
\end{equation}
\vspace{-0.14in}
\begin{eqnarray}
	\overline I(i, j_{\beta} -m < j \leq j_\beta) = \frac{1}{m}\sum_{ j_\beta -m < j \leq j_\beta} I(i, j),
\end{eqnarray}
where $i$ and $j_\beta$ are indices of a pixel in an unfolded frame $I$, $m$ is a hyper-parameter marking the range, and $\overline I$ denotes calculating the mean value. As shown in Fig.~\ref{diff}, the measure $M_I$ is capable of distinguishing FA and non-FA areas.

\begin{figure}[t]
\centerline{\includegraphics[width=0.7\columnwidth]{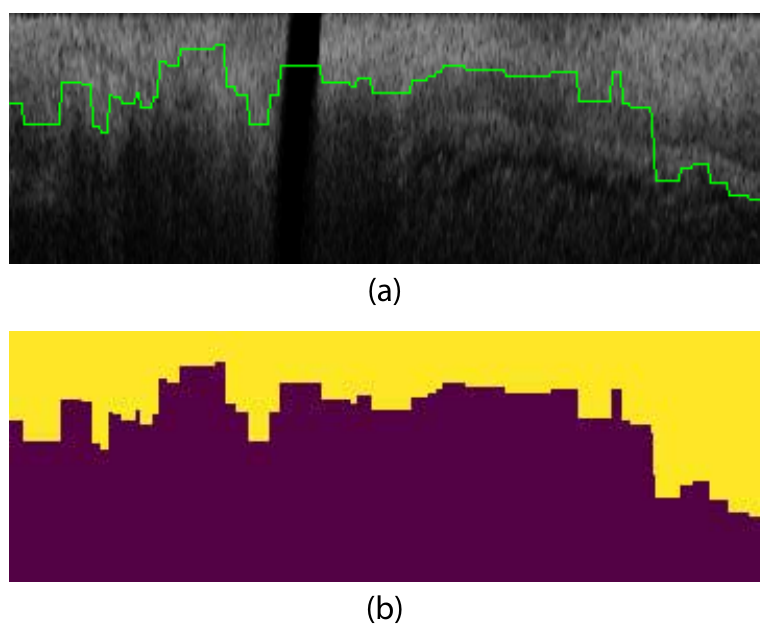}}
\caption{Illustrating (a) a masked frame and (b) the \textcolor{orange}{orange} mask between the lumen-intima (LI) and adventitia–periadventitia (AP) surfaces.}

\label{border}
\end{figure}

\noindent
\textbf{Binary Mask Image (BMI):} In addition to the above two types of features, the distances between the lumen-intima (LI) and adventitia–periadventitia (AP) surfaces can also help FA identification. As shown in Fig.~\ref{border}, an area tends to exhibit different cap brightness and shadows between the LI and AP surfaces. Using the dynamic programming method in \citep{Zahnd}, we first detect the AP surface (the green curve in Fig.~\ref{border}(a)), then convert the frame into a binary mask (shown in Fig.~\ref{border}(b)), and utilize it to extract FA features. Specifically, given an original frame $OF$ of size $H \times W$, the AP surface detected by the method in~\citep{Zahnd} is a vector $[s_1, s_2, \ldots, s_{W}]$, where $s_i$ is the position of the AP surface at the $i$-th column.  We first define a zero matrix $M$ of size $H\times W$, and then build $M$ as:
\begin{eqnarray}
    M[0:s_i, i] = 1, \quad\quad i = 1, 2, \ldots, W,
\end{eqnarray}
where $M$ is a binary mask image (BMI, shown in Fig.~\ref{border}(b)). 

With an original frame $I$ and its three different auxiliary images, which capture intensity changes along the radial direction and the thickness between the LI and AP surfaces, we develop a multi-encoder segmentation approach to obtain the corresponding features. With four binary partition trees (BPTs) of the same structure, we take $I$ and its auxiliary images as input, and compute their feature maps. Instead of concatenating these feature maps, we apply a multi-head cross-attention mechanism~\citep{vaswani2017attention} to fuse them, as shown in Fig.~\ref{BinaryTree}(a).

Specifically, consider a root-to-leaf path $p$ in a BPT for a frame $F\in \{OF, GI, LGI, BMI\}$. We denote the output feature map for (a sub-region of) $F$ at level $i$ of $p$ as $f'_{i}(F)$ (see Eq.~(\ref{aggregated-feature-map})), where $f'_{i}(F)\in R^{C\times H\times W_i}$, $C, H,$ and $W_i$ are for the feature map channels, height, and width of the sub-region of $F$ at the $i$-th level of $p$ respectively, $W_i=\frac{W_0}{2^i}$, and $W_0$ is $F$'s width. We perform an average pooling on $f'_{i}(F)$ and obtain a feature vector  $l_{i}(F)\in R^C$. We use $l_{i}(OF)$ as query and $l_{i}(OF), l_{i}(GI), l_{i}(LGI)$, and $l_{i}(BMI)$ as keys, and define three matrices $W_q, W_k$, and $W_v$, which contain learnable parameters and are randomly initialized. The mapped query $Q_i$, key $K_i$, and value $V_i$ matrices are computed as:
\begin{eqnarray}
Q_i(F) = l_i(F)\times W_q, \\
K_i(F) = l_i(F)\times W_k, \\
V_i(F) = W_v(f'_i(F)),
\end{eqnarray}
where $W_q, W_k\in R^{C\times C}$, $W_v$ is $1\times 1$ convolution, and $V_i(F)$ is a mapped feature map matrix, with $V_i(F)\in R^{C\times H\times W_i}$. The final, aggregated feature map at level $i$ is 
computed as:
\begin{equation}
f''_i = \sum_{F\in\{OF, GI, LGI, BMI\}}\phi(\frac{Q_i(OF)*K_i(F)^T}{\sqrt{C}}) V_i(F),
\end{equation}
where $\phi$ is a softmax function.

\noindent
\textbf{Loss Function}: The overall loss on the path $p$ is:
\begin{eqnarray}
	{\cal {L}} = \sum_{i=0}^{m-1} {\cal {L}}_i(f''_i, GT_i),
\end{eqnarray}
where ${\cal {L}}_i$ is the cross-entropy loss of the $i$-th level aggregated feature map $f''_i$ and the ground truth $GT_i$. 

\section{Experiments}
\subsection{Dataset}
We searched for public datasets for this study, but could not find any.\footnote{We contacted the authors of~\cite{lee2024deep} and were told that their data could not be shared due to privacy issues with a hospital.} 
The dataset we used was collected from 
56 patients with symptomatic stable CAD  as part of the Charles University in Prague ``Prediction of Extent and Risk Profile of Coronary Atherosclerosis and Their Changes During Lipid-lowering Therapy Based on Non-invasive Techniques'' (PREDICT) trial (NCT01773512). Enrolled patients underwent angiography and culprit lesion percutaneous coronary intervention. A subset of 24 patients from this cohort who underwent OCT imaging at the time of baseline procedure and 12-month follow-up. OCT imaging was performed in the identical vessel segment via a frequency-domain ILUMIENS OCT catheter (St.~Jude Medical). After contrast administration via power injection to create a blood-free lumen, OCT images were recorded at 20~mm/s for a total length of 54 mm. 
Voxel spacing of the OCT pullback was $0.0149$$\times$$0.0149$$\times$$0.1990$ (mm). Each frame of a 3D stack was annotated with a value of 0 or 1 for each the 360 degrees circumferentially, indicating whether the ray at that angle anchoring at the frame center contained FA or not.

\begin{figure*}[!ht]
    \centering    
    \includegraphics[width=1.2\columnwidth]{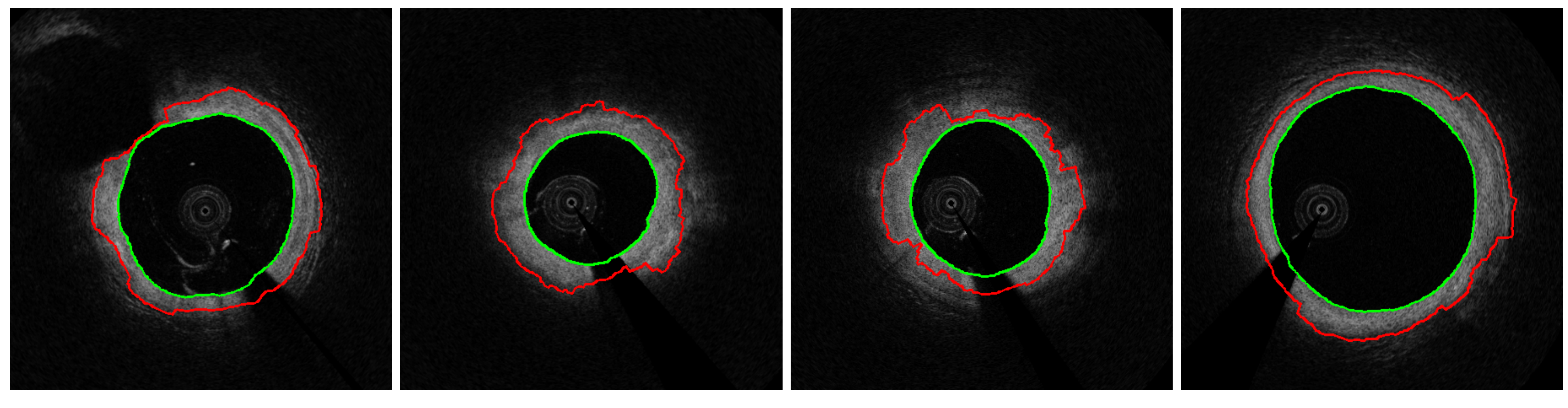}
    \caption{Qualitative results of detecting the lumen-intima (LI, \textcolor{green}{green}) and adventitia-periadventitia (AP, \textcolor{red}{red}) surfaces.}
    \label{fig:border_result}
\end{figure*}

\begin{figure}	
\centerline{\includegraphics[width=0.65\columnwidth]{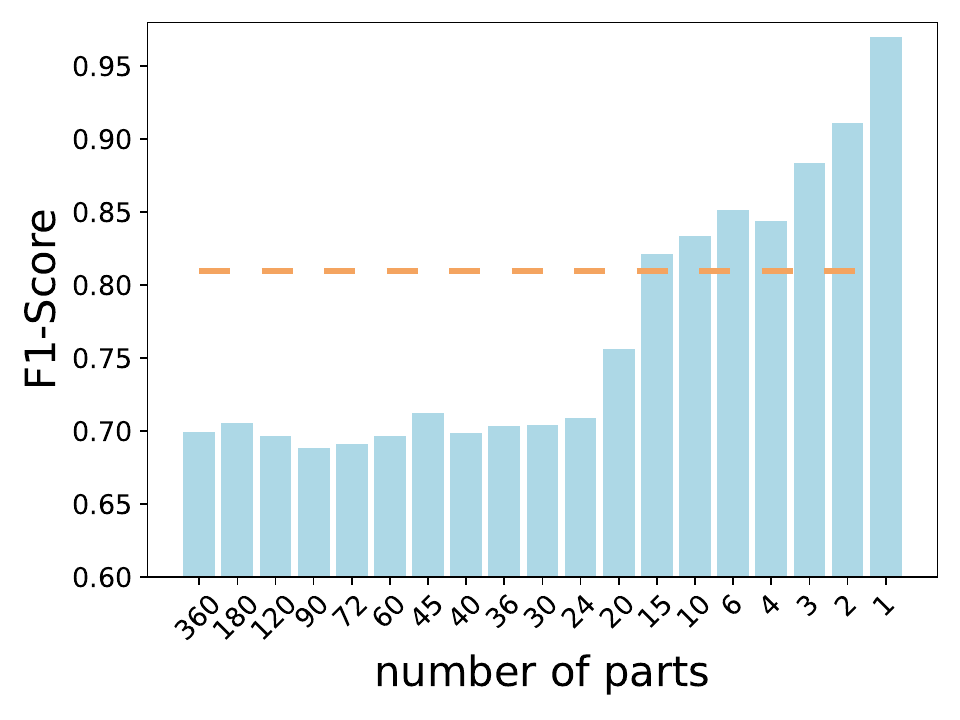}}
\caption{The F1-scores for dividing a frame into different numbers of parts. The orange dashed line indicates our final results in Table \ref{table:main_result}.
}
\label{fig:performance_ratio}
\end{figure}

\begin{figure*}[h!]
\centering
\includegraphics[width=2.0\columnwidth]{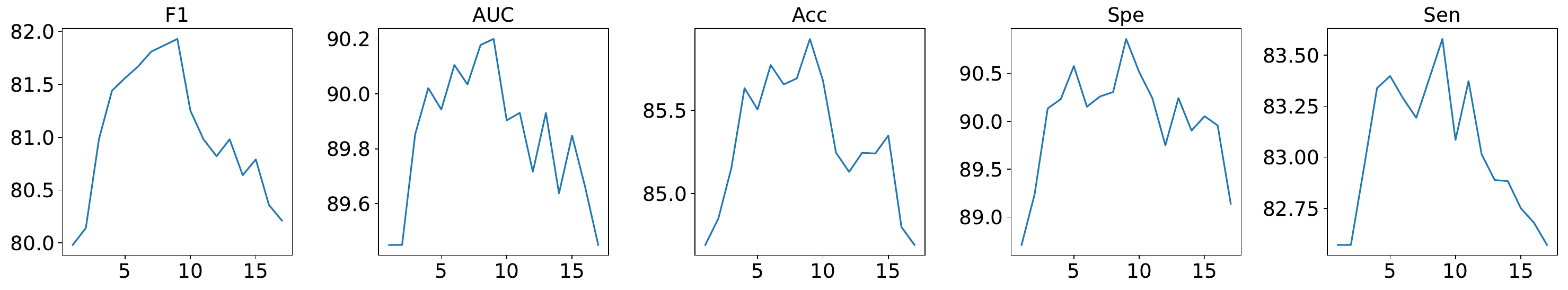}

\caption{Ablation study of the range value selection in generating long-range gradient images.}
\label{fig:range}
\end{figure*}

\begin{figure*}[!ht]                                   
\centerline{\includegraphics[width=1.06\columnwidth]{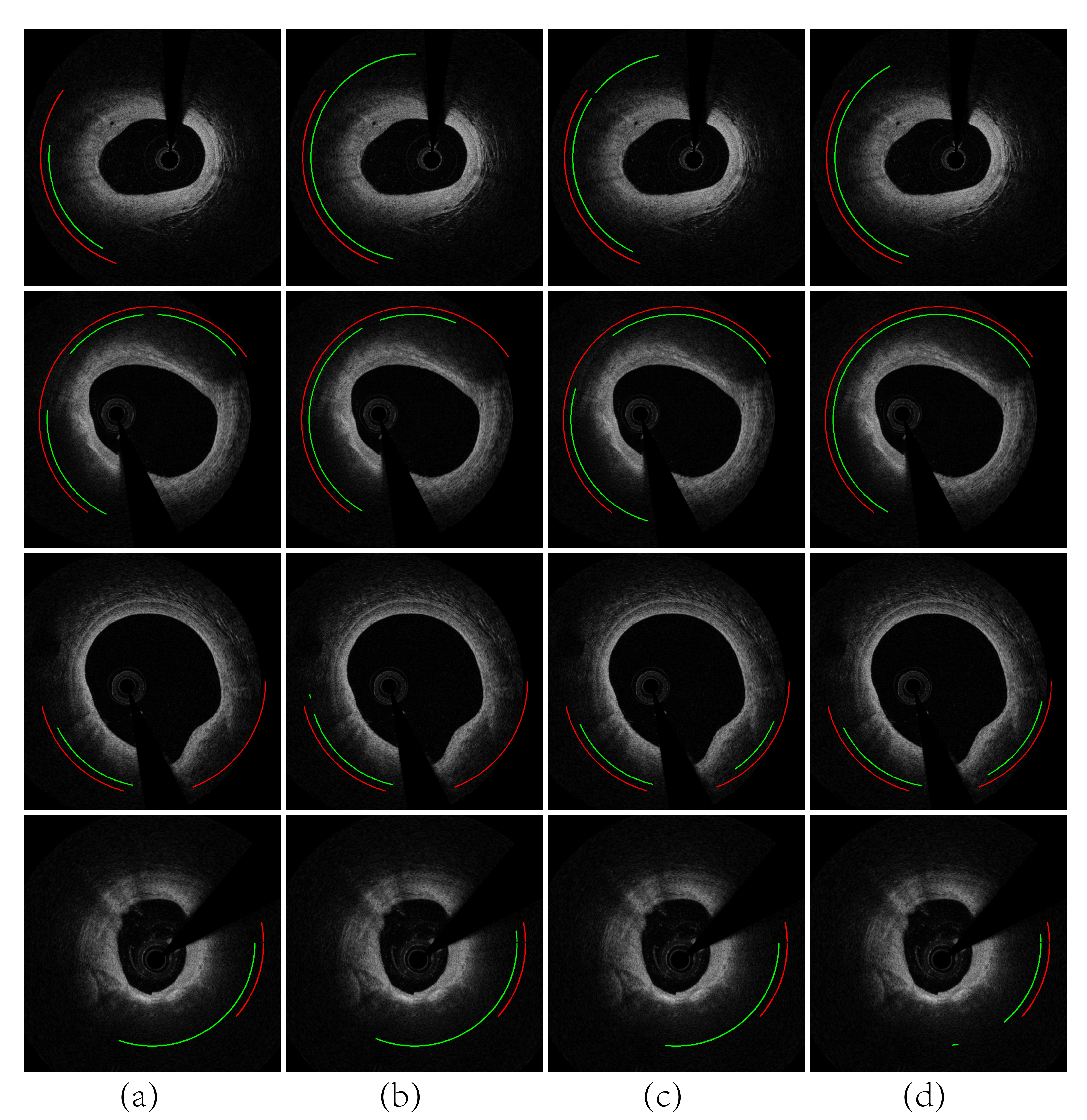}}
\caption{
Examples of qualitative results by different methods for the FA range detection problem. We compare our approach with three known methods which are the top three in Table~\ref{table:main_result}. Columns (a)-(d) give results of Swin-Unet~\citep{cao2023swin}, G-Swin-Transformer~\citep{wang2023vision}
Transfer-OCT~\citep{lee2024deep}, 
and our FiAt-Net, respectively. Colors {\color{red}{red}} and {\color{green}{green}} indicate the ground truth and results generated by different methods, respectively.}
\label{qua_res}
\end{figure*}

\subsection{Implementation Details}

We implement our FiAt-Net model using PyTorch~\citep{paszke2019pytorch}. We select Swin-Unet~\citep{cao2023swin} as our backbone since it outperforms other methods (see Table~\ref{vanilla_seg}). The initial learning rate is set to 0.0001. We train the model for 1000 epochs and use polynomial learning rate decay with a power of 0.9 to smooth the training process. All the networks are trained on NVIDIA Tesla P100s (16GB VRAM). In the FA search process, if two closest, isolated {\it positive} angles detected are $\leq 4$ degrees apart, we consider them as noise and treat them as {\it negative}. We empirically set the 
threshold $\alpha$ = 4, meaning that a sub-region $R$ is taken as a negative sample if its non-FA proportion exceeds $L-4$, where $L$ is the angle range size of $R$. The dimension of the feature map $f$ used to compute the attention score and the LGI range parameter $m$ are empirically set to 1024 and 9, respectively.

In the FA search process, we use 
a DFS-like  scheme, as: If the positive portion is $<4$\textdegree~on a patch, we stop the partition on the patch; else, we split the patch into two parts and search each part. At each lower level, the proportion of positive areas will increase, thus reducing the imbalanced effect.

\subsection{Evaluation Metrics}
We use five common evaluation metrics: F1-score, accuracy, area under the receiver operating characteristic (ROC) curve (AUC), sensitivity, and specificity. 
{\bf Regarding the fibrous cap measurements, it is typical to mark a single point or several discrete points on the fibrous cap to measure its thickness. As such, the available annotations are A-line based. It appears that A-line based information and angular information are equally important, as they indicate the angular range of the potential lipid pool behind the fibrous cap. However, due to the absence of angular marking functionality in routine clinical tools, angular extent information is often not available in clinical practice.} 

Both F1-score and AUC evaluate the overall performance of a model. F1-score focuses more on the positive areas; a higher F1-score indicates that the model is capable of detecting FA areas effectively. AUC, on the other hand, focuses more on the negative areas, suggesting that the model pays more attention to non-FA regions. Specificity measures the proportion of successfully detected negative samples, while sensitivity measures the proportion of successfully detected positive regions.

\subsection{Experimental Setup}
In the experiments, we use Swin-Unet \citep{cao2023swin} as our backbone. We compare our FiAt-Net approach with the following known methods: (1) U-Net \citep{ronneberger2015u}; (2) TransUNet \citep{chen2102transformers}; (3) PraNet \citep{fan2020pranet}; (4) Attention U-Net \citep{oktay2018attention}; (5) Zahand et al.'s method \citep{zahnd2017contour}; (6) DomainNet \citep{shi2018vulnerable}; (7) G-Swin-Transformer~\citep{wang2023vision}; (8) Transfer-OCT~\citep{lee2024deep}; (9) Swin-Unet \citep{cao2023swin}. We also compare with two typical data imbalance mitigating methods: (10) focal loss \citep{lin2017focal}, which is a loss function focusing on learning from hard examples; (11) class imbalance loss \citep{cui2019class}, a re-weighting scheme that uses an effective number of samples for each class to re-balance the loss. All the experiments are evaluated using five-fold cross-validation, in which 5 patients are used for testing and 19 patients are used for training.

\subsection{Experimental Results}
In the experiments, we use the dynamic programming method in~\citep{Zahnd} to segment the lumen-intima (LI) and adventitia-periadventitia (AP) surfaces, for which ground truth labels are not provided. We present representative  segmentation results in Fig.~\ref{fig:border_result}. From Fig.~\ref{fig:border_result}, we observe that the LI surface detection is quite accurate because the border between the lumen and intima is fairly clear. Although some artifacts may affect the AP surface detection results, their influence on the final FA detection is quite limited since this is only one piece of information fed to the model.

The main results and comparisons are shown in Table~\ref{table:main_result}. One can see that our pre-processing and clustering methods are able to help boost the performance in all the evaluation metrics. Our FiAt-Net approach outperforms state-of-the-art methods. For example, compared to focal loss~\citep{lin2017focal} and class imbalance loss~\citep{cui2019class}, our approach improves the performance by over 7.0\% in F1-score. The improvement is attributed to our proposed binary partition tree (BPT), which filters out sub-regions that do not contain 
FA at a high level. Consequently, the models are trained with more FA-containing regions, thus mitigating the sparse distribution of FA areas. The ablation study on BPT in Section~\ref{bpt_ablation} demonstrates the effectiveness of the BPT mechanism. 
The improvement in AUC and accuracy is limited because the negative areas dominate the distribution of the dataset. Thus, the improvement in positive area detection brings only a limited performance gain in these two evaluation metrics. Table~\ref{table:main_result} also demonstrates that the imbalanced data processing techniques can significantly improve the accuracy of positive FA area detection in some of these metrics.

\begin{table*}[!ht]
\caption{Comparisons of experimental results obtained by different methods. The pre-processing and clustering methods are applied to all the experiments except for the first two rows.}
\label{table:main_result}
\centering
\begin{tabu}{r|c|c|c|c|c}
Method & F1 & AUC & Acc & Spe & Sen  \\

\hline
Swin-Unet~\citep{cao2023swin} w/o Pre-processing & 57.69$\pm$3.68 & 78.96$\pm$3.45& 78.98$\pm$2.53&81.36$\pm$2.99 &69.67$\pm$3.26 \\
Swin-Unet~\citep{cao2023swin} w/o Clustering & 67.84$\pm$2.42 &83.98$\pm$3.02 & 81.26$\pm$3.47 & 84.03$\pm$3.49 &72.19$\pm$1.68 \\
U-Net~\citep{ronneberger2015u} & 63.45$\pm$2.13 & 79.81$\pm$1.91 & 80.72$\pm$2.54 & 84.34$\pm$2.19&75.31$\pm$2.78 \\
TransUNet~\citep{chen2102transformers} & 63.82$\pm$3.12 & 83.07$\pm$2.18& 81.32$\pm$1.27 &85.62$\pm$2.34& 78.02$\pm$2.47  \\
PraNet~\citep{fan2020pranet} & 65.19$\pm$2.41 & 81.37$\pm$2.68 & 81.29$\pm$2.45 &86.01$\pm$1.69 &78.94$\pm$2.51\\

Attention U-Net~\citep{oktay2018attention} & 66.97$\pm$2.54 & 85.77$\pm$1.57& 82.05$\pm$2.39 & 85.49$\pm$2.03 & 75.99$\pm$2.47\\

Zahnd et al.\citep{zahnd2017contour} & 63.28$\pm$2.34 & 79.11$\pm$1.99 & 79.50$\pm$2.63  & 83.05$\pm$2.64 & 73.54$\pm$2.58\\

DomainNet~\citep{shi2018vulnerable} & 65.24$\pm$2.04 & 81.79$\pm$2.36 & 80.22$\pm$2.97 & 82.94$\pm$3.04 & 72.69$\pm$2.78\\

G-Swin-Transformer~\citep{wang2023vision} & 74.98$\pm$2.57 & 90.32$\pm$3.04 & 83.00$\pm$2.08 & 86.54$\pm$3.59 & 82.59$\pm$3.33\\

Transfer-OCT~\citep{lee2024deep} & 76.96$\pm2.47$ & 89.43$\pm$2.12 &  83.33$\pm$2.94 & 88.43$\pm$3.74 & 81.92$\pm$3.09 \\

Swin-Unet~\citep{cao2023swin} & 69.95$\pm$2.44 & 88.80$\pm$2.35 & 83.45$\pm$1.69 & 87.19$\pm$2.41 & 78.07$\pm$2.63 \\

Swin-Unet~\citep{cao2023swin} +  & \multirow{2}{*}{71.32$\pm$2.74} & \multirow{2}{*}{90.23$\pm$2.47} & \multirow{2}{*}{82.99$\pm$2.54} & \multirow{2}{*}{88.02$\pm$1.99} &\multirow{2}{*}{82.00$\pm$2.13}\\		

Focal Loss~\citep{lin2017focal} \\

Swin-Unet~\citep{cao2023swin} +  & \multirow{2}{*}{74.32$\pm$2.08} & \multirow{2}{*}{\textbf{90.64$\pm$2.67}}&\multirow{2}{*}{82.36$\pm$1.67} & \multirow{2}{*}{87.96$\pm$2.97} & \multirow{2}{*}{82.27$\pm$1.69}\\
Class Imbalance Loss~\citep{cui2019class}
\\

\hline 
		
FiAt-Net (ours) & \textbf{81.93$\pm$2.50} &  90.20$\pm$2.34 & \textbf{85.93$\pm$1.96} & \textbf{90.86$\pm$3.08} & \textbf{83.58$\pm$2.44}
		
\end{tabu}

\end{table*}

\begin{table*}[]
\centering
\caption{Ablation study of the frame clustering mechanism on different FA detection methods.}
\begin{tabular}{r|c|c|c|c|c|c}  
Method & Frame Clustering & F1 & AUC & Acc & Spe & Sen \\
\hline

\multirow{2}{*}{Swin-UNet~\citep{cao2023swin}} & \xmark & 67.84$\pm$2.42 & 83.98$\pm$3.02 & 81.26$\pm$3.47 & 84.03$\pm$3.49 &72.19$\pm$1.68 \\
 & \checkmark & 69.95$\pm$2.44 & 88.80$\pm$2.35 & 83.45$\pm$1.69 & 87.19$\pm$2.41 & 78.07$\pm$2.60\\
 
\hline
\multirow{2}{*}{G-Swin-Transformer~\citep{wang2023vision}} & \xmark & 74.98$\pm$2.57 & 85.32$\pm$3.04 & 82.00$\pm$2.08 & 86.54$\pm$3.59 & 78.85$\pm$3.33\\

& \checkmark &  77.69$\pm$2.23 & 87.01$\pm$2.81 & 84.23$\pm$1.90 & 87.99$\pm$3.20 & 80.72$\pm$3.54\\

 \hline
\multirow{2}{*}{Transfer-OCT~\citep{lee2024deep}} &\xmark &  76.96$\pm2.47$ & 86.43$\pm$2.12 &  82.33$\pm$2.94 & 85.43$\pm$3.74 & 79.92$\pm$3.09\\
 
 & \checkmark & 78.59$\pm$2.92 & 88.76$\pm$1.95 &  84.01$\pm$3.17 & 86.24$\pm$3.20 & 81.90$\pm$2.59\\

\hline
\multirow{2}{*}{FiAt-Net (ours)} & \xmark & 80.07$\pm$2.41 &  89.36$\pm$2.05 & 85.00$\pm$2.13 & 89.91$\pm$2.84 & 82.85$\pm$2.90\\

& \checkmark & \textbf{81.93$\pm$2.50} &  \textbf{90.20$\pm$2.34} & \textbf{85.93$\pm$1.96} & \textbf{90.86$\pm$3.08} & \textbf{83.58$\pm$2.44}\\

\end{tabular}

\label{tab:clustering_1}
\end{table*}

\begin{table*}[]
\centering

\caption{Ablation study of different methods for detecting the Lumen-Intima (LI) border.}
\begin{tabular}{r|c|c|c|c|c}
Method & F1 & AUC & Acc & Spe & Sen  \\
\hline
FiAt-Net w/o pre-processing & 76.21$\pm$2.98 &  86.31$\pm$2.74 & 82.15$\pm$2.74 & 86.45$\pm$3.19 & 80.18$\pm$2.97 \\

FiAt-Net + \citep{shi2023automatic} & 81.78$\pm$2.44 &  \textbf{90.29$\pm$2.51} & 84.79$\pm$2.00 & 90.68$\pm$3.45 & 83.21$\pm$2.76 \\

FiAt-Net + \citep{chen2023deep} & \textbf{82.06$\pm$1.98} &  90.02$\pm$2.09 & \textbf{86.15$\pm$1.78} & 90.24$\pm$3.52 & 83.41$\pm$2.69 \\

FiAt-Net + \citep{zahnd2017contour} & 81.93$\pm$2.50 &  90.20$\pm$2.34 & 85.93$\pm$1.96 & \textbf{90.86$\pm$3.08} & \textbf{83.58$\pm$2.44} 

\end{tabular}

\label{tab:LI_method}
\end{table*}

\begin{table*}[!ht]
\centering
\caption{Experimental results of ablation study for binary partition and auxiliary images. BP, GI, WGI, and BMI denote binary partition, Gradient Image, Wide-range Gradient Image, and Binary Mask Image, respectively.}
\label{table:diff_image}
\begin{tabular}{cccc|c|c|c|c|c}
\multicolumn{4}{c|}{Setting} & \multirow{2}{*}{F1} & \multirow{2}{*}{AUC} & \multirow{2}{*}{Acc} & \multirow{2}{*}{Spe} & \multirow{2}{*}{Sen}  \\

\cline{1-4}
BP & GI & WGI & BMI & & & & & \\
\hline
\checkmark & \checkmark & & & 77.45$\pm$1.69 & 88.81$\pm$2.35 & 83.72$\pm$3.45 &89.12$\pm$1.94 & 79.65$\pm$2.79\\

\checkmark & & \checkmark & & 77.82$\pm$2.45 & 89.07$\pm$3.05& 84.32$\pm$2.19 & 88.94$\pm$2.36& 81.69$\pm$1.95 \\

\checkmark & & & \checkmark &
 76.69$\pm$1.99 & 88.37$\pm$2.48 & 83.29$\pm$2.57 &88.06$\pm$2.64 &81.03$\pm$2.58\\

\checkmark & \checkmark & \checkmark & & 79.36$\pm$2.23& 89.36$\pm$3.02 & 84.25$\pm$3.01 &89.24$\pm$2.45 & 82.08$\pm$3.21\\

\checkmark & \checkmark & & \checkmark &79.98$\pm$1.86 & 89.45$\pm$2.16&84.69$\pm$2.29&88.71$\pm$3.27&82.57$\pm$2.99\\

\checkmark & & \checkmark & \checkmark & 80.15$\pm$2.61 & 88.98$\pm$2.69&85.01$\pm$1.79&88.43$\pm$2.78&83.05$\pm$1.76\\

& \checkmark  & \checkmark & \checkmark & 71.24$\pm$2.30 & 88.95$\pm$1.87 & 84.01$\pm$3.14 &87.99$\pm$2.10 &81.31$\pm$1.99\\

\checkmark & \checkmark  & \checkmark & \checkmark & \textbf{81.93$\pm$2.50}&  \textbf{90.20$\pm$2.34} &\textbf{85.93$\pm$1.96} & \textbf{90.86$\pm$3.08} & \textbf{83.58$\pm$2.44} 
		
\end{tabular}
\end{table*}

\begin{table*}[!ht]
\centering
\caption{Experimental results for using different feature fusion methods. SE, CFF, ME, and AFF denote shared encoder, concatenation feature fusion, multiple encoders, and attention feature fusion, respectively.}
\label{table:fusion}
\begin{tabular}{cccc|c|c|c|c|c}
\multicolumn{4}{c|}{Feature Fusion Method} & \multirow{2}{*}{F1} & \multirow{2}{*}{AUC} & \multirow{2}{*}{Acc} & \multirow{2}{*}{Spe} & \multirow{2}{*}{Sen}  \\

\cline{1-4}
SE & CFF & ME & AFF & & & & \\

\hline

\checkmark & \checkmark & & & 77.45$\pm$2.64 & 89.02$\pm$1.81 & 84.72$\pm$2.54 & 88.29$\pm$1.89 & 80.35$\pm$3.45 \\

\checkmark & & & \checkmark & 80.28$\pm$1.94 & 89.32$\pm$2.94& 85.33$\pm$2.39 & 88.33$\pm$1.98 & 81.99$\pm$2.97 \\

& \checkmark & \checkmark & & 79.79$\pm$3.01 & 89.53$\pm$2.34 & 84.99$\pm$2.69 & 89.24$\pm$2.64 & 81.56$\pm$1.95\\

& & \checkmark & \checkmark & \textbf{81.93$\pm$2.50} &  \textbf{90.20$\pm$2.34} & \textbf{85.93$\pm$1.96} & \textbf{90.86$\pm$3.08} & \textbf{83.58$\pm$2.44} 
		
\end{tabular}
\end{table*}

\subsection{Ablation Studies}
We present ablation studies 
to examine the effectiveness of each our key proposed component (not to be confused with the coronary ablation procedures in interventional cardiology).

\noindent
{\bf Ablation Study of Frame Clustering}: We evaluate the effectiveness of our proposed frame clustering method by testing it on four methods: 1) Swin-UNet~\citep{cao2023swin}; 2) G-Swin-Transformer~\citep{wang2023vision}; 3) Transfer-OCT~\citep{lee2024deep}; 4) our FiAt-Net. The experimental results are given in Table~\ref{tab:clustering_1}. From these results, we observe that our frame clustering mechanism consistently improves FA detection scores in terms of F1, AUC, Acc, Spe, and Sen, because the clustering mechanism allows different types of plaques to be uniformly sampled, thus mitigating the imbalanced distribution issue.

\noindent
\label{li_process}
{\bf Ablation Study of Different Lumen-Intima (LI) Border Detection Methods:} We examine several methods for detecting the LI border. Specifically, we test three typical methods: 1)~\citep{zahnd2017contour}'s method; 2) \citep{shi2023automatic}'s method; 3) \citep{chen2023deep}'s method. The experimental results are shown in Table~\ref{tab:LI_method}. As observed from Table~\ref{tab:LI_method}, the F1-score drops by 5.72\% when the pre-processing step is not applied. This further demonstrates that the lumen background (including probe, blood remnants, etc.) can distract the model and decrease performance. Note that the differences between different Lumen-Intima border detection methods are quite limited.

\noindent
{\bf Ablation Study of Long-range Gradient Images:} For the auxiliary images, the performance may be sensitive to the range value $m$ of the long-range gradient images. Hence, we conduct experiments on range value selection, as shown in Fig.~\ref{fig:range}. From Fig.~\ref{fig:range}, the performance increases with the increase in the range, and starts to decline when the range exceeds 9.

\noindent
{\bf Ablation Study of the Binary Partition Method:}
We conduct ablation experiments to examine the effect of our binary partition method by dividing an input frame into various numbers of parts.
Note that our main task is to detect FA ranges in 360 angles. However, directly detecting FA in each of 360 individual angles in a frame does not yield good results (the F1-score of dividing a frame into 360 parts in Fig.~\ref{fig:performance_ratio} is only $\sim$70\%). Thus, we consider a simplified case: determine whether a frame/sub-region contains FA without specifying the specific FA locations in it. This is actually a classification problem. 
When treating the frame as a whole (i.e., with only one part), the performance of this simplified case can exceed 95\%. But, this simplified case does not solve our task, as we require output of specific FA locations. To take further steps, we divide the frame equally into two sub-regions and decide whether each sub-region contains FA (with 2 parts in Fig.~\ref{fig:performance_ratio}); the F1-score drops to $\sim$90\%. As we go into finer scales, the performance drops more. When dividing the frame into 360 parts, the performance is only $\sim$70\%. 
Based on these experiments, our FiAt-Net applies a binary partition method by incorporating different scales to improve the performance from the finest scale 
(from $\sim$70\% to $\sim$82\%, indicated by the \textcolor{orange}{orange} dashed line in Fig.~\ref{fig:performance_ratio}).

\noindent
\label{bpt_ablation}
{\bf Ablation Study of the Auxiliary Images:}
We examine the effectiveness of our binary partition and auxiliary images. 
As shown in Table~\ref{table:diff_image}, 
binary partition and these three types of auxiliary images help
improve the FA range detection performance by $>4\%$ in F1-score. This is because the density distribution reveals informative clues on whether FA exists.

\noindent
\label{bpt_ablation}
{\bf Ablation Study of Different Fusion Methods:}
There are various ways to fuse multiple images (possibly of different types). A simple way is to concatenate them as a multi-channel image, and feed it to an encoder. We report experimental results of several fusion methods in Table~\ref{table:fusion}. From Table~\ref{table:fusion}, one can see that using multiple encoders to process the original frame, Gradient Image, Wide-range Gradient Image, and Binary Mask Image can improve performance. These images have different appearances and present different information. Using multiple encoders allows the model to focus on the critical information presented in each of these images. Compared to concatenation fusion, the attention-based fusion mechanism enhances performance since not all features are of equal importance. The attention fusion mechanism assigns a learnable weight to each kind of features, allowing the model to put different emphases on different features.


		

\subsection{Qualitative Results}
We show some qualitative results in Fig.~\ref{qua_res}, in which columns (a)-(d) give the results of Swin-Unet \citep{cao2023swin}, 
G-Swin-Transformer~\citep{wang2023vision}, Transfer-OCT~\citep{lee2024deep}, 
and our FiAt-Net, respectively. Note that we detect FA angular ranges in the polar domain, and project them back to the Cartesian domain. From Fig.~\ref{qua_res}, we observe that even though the annotations may be noisy, our FiAt-Net is still able to capture FA ranges by excluding the guide-wire shadow areas. Also, our approach can detect some small FA areas, which could be missed by the other methods.


\subsection{Parameters and Computational Complexity}

In Table~\ref{complexity}, we report the parameters and computational complexity of our FiAt-Net and Swin-Unet~\citep{cao2023swin} on an NVIDIA P100 GPU. The higher parameter and computation costs of our FiAt-Net are primarily introduced by the multi-head encoder, and are within a reasonable range. The performance improvement that we gain suggests that the additional computation costs, based on Swin-Unet \citep{cao2023swin}, are worthwhile.

\setlength{\tabcolsep}{1pt}
\begin{table}
\centering
\caption{Comparison of parameters and computational complexity (on an NVIDIA P100 GPU).
}
\label{complexity}
\begin{tabular}{c|c|c|c|c}
Method & F1-Score & \# Params. & FLOPs & FPS \\		
\hline 				
Swin-Unet~\citep{cao2023swin} & 69.95 & 27.17M & 17.49G & 22.32 \\
FiAt-Net (ours) & 81.93 & 51.68M & 28.00G &  16.70 \\
\end{tabular}
\end{table}

\subsection{Discussions}

We distinguish ourselves from the known methods by introducing an auxiliary image representation and applying the attention mechanism to fuse features of the input and auxiliary images. The experimental results in Table~\ref{table:main_result} demonstrate the 
advantages of our approach over the known methods. Furthermore, the ablation study of our proposed auxiliary images (\textit{BP, GI, WGI, and BMI}) in Table~\ref{table:diff_image} shows the effectiveness of each auxiliary image. The ablation study of our proposed attention feature fusion in Table~\ref{table:fusion} also demonstrates the effectiveness of our attention feature fusion (AFF) method. All these show that our approach is capable of utilizing the characteristics of different auxiliary images and effectively fusing their features.

\noindent
{\bf The Effect of Sparse Occurrences of FA:} 
From Fig.~\ref{fig:performance_ratio},
we observe that a large number of parts will hinder the model's ability to learn representations and lead to a decrease in F1-score. Thus, we propose to use binary partition to gradually narrow down the FA areas, amplifying the model's focus on FA regions while attenuating the influence of non-FA regions. As shown in Table~\ref{table:diff_image},
we see that the binary partition scheme helps improve the F1-score by more than 10\%, further demonstrating that sparse occurrences of FA can hinder DL model performance, and our method can mitigate such issues.

\noindent
{\bf Advantages Compared to Known IVOCT Analysis Methods:} The core advantages of our proposed approach 
result from incorporating: 1) a pre-processing step to remove some background (including probe, blood remnants, etc.); 2) a frame clustering mechanism, enabling different types of plaques to be uniformly sampled, thus mitigating the issue of imbalanced plaque distribution; 3) a binary partition mechanism to gradually narrow down the FA areas, amplifying the model's focus on FA regions while attenuating the influence of non-FA regions; and 4)~improved performance compared to previous methods~\citep{zahnd2017contour, shi2018vulnerable, lee2024deep, wang2023vision}.

\section{Conclusions}
In this paper, we proposed a new approach, FiAt-Net, for detecting the cap of fibroatheroma (FA) in 3D IVOCT images. 
We applied a frame clustering method and sampled 2D frames from each cluster, making the data in training batches close to the distribution of the dataset. We presented a binary partition scheme to progressively narrow down the FA areas. We constructed additional image representations 
(auxiliary images) to help distinguish FA and non-FA areas. 
We developed a multi-head encoder to incorporate diverse information, and applied an attention fusion mechanism to fuse multi-level features from the original and auxiliary images. Extensive experiments and ablation study demonstrated the effectiveness of our new approach for FA detection.

\vspace{0.1in}
\noindent
{\bf Acknowledgment.}
This research was supported in part by NIH Grant 2R56EB004640-16.

\bibliographystyle{model2-names.bst}\biboptions{authoryear}
\bibliography{refs}

\begin{thebibliography}{34}
\expandafter\ifx\csname natexlab\endcsname\relax\def\natexlab#1{#1}\fi
\providecommand{\url}[1]{\texttt{#1}}
\providecommand{\href}[2]{#2}
\providecommand{\path}[1]{#1}
\providecommand{\DOIprefix}{doi:}
\providecommand{\ArXivprefix}{arXiv:}
\providecommand{\URLprefix}{URL: }
\providecommand{\Pubmedprefix}{pmid:}
\providecommand{\doi}[1]{\href{http://dx.doi.org/#1}{\path{#1}}}
\providecommand{\Pubmed}[1]{\href{pmid:#1}{\path{#1}}}
\providecommand{\bibinfo}[2]{#2}
\ifx\xfnm\relax \def\xfnm[#1]{\unskip,\space#1}\fi
\bibitem[{Abdolmanafi et~al.(2020)Abdolmanafi, Cheriet, Duong, Ibrahim and Dahdah}]{Abdolmanafi}
\bibinfo{author}{Abdolmanafi, A.}, \bibinfo{author}{Cheriet, F.}, \bibinfo{author}{Duong, L.}, \bibinfo{author}{Ibrahim, R.}, \bibinfo{author}{Dahdah, N.}, \bibinfo{year}{2020}.
\newblock \bibinfo{title}{An automatic diagnostic system of coronary artery lesions in {Kawasaki} disease using intravascular optical coherence tomography imaging}.
\newblock \bibinfo{journal}{Journal of Biophotonics} \bibinfo{volume}{13}, \bibinfo{pages}{e201900112}.
\bibitem[{Cao et~al.(2023)Cao, Wang, Chen, Jiang, Zhang, Tian and Wang}]{cao2023swin}
\bibinfo{author}{Cao, H.}, \bibinfo{author}{Wang, Y.}, \bibinfo{author}{Chen, J.}, \bibinfo{author}{Jiang, D.}, \bibinfo{author}{Zhang, X.}, \bibinfo{author}{Tian, Q.}, \bibinfo{author}{Wang, M.}, \bibinfo{year}{2023}.
\newblock \bibinfo{title}{{Swin-Unet}: Unet-like pure {Transformer} for medical image segmentation}, in: \bibinfo{booktitle}{ECCV, Part III}, pp. \bibinfo{pages}{205--218}.
\bibitem[{Chen et~al.(2021)Chen, Lu, Yu, Luo, Adeli, Wang, Lu, Yuille and Zhou}]{chen2102transformers}
\bibinfo{author}{Chen, J.}, \bibinfo{author}{Lu, Y.}, \bibinfo{author}{Yu, Q.}, \bibinfo{author}{Luo, X.}, \bibinfo{author}{Adeli, E.}, \bibinfo{author}{Wang, Y.}, \bibinfo{author}{Lu, L.}, \bibinfo{author}{Yuille, A.L.}, \bibinfo{author}{Zhou, Y.}, \bibinfo{year}{2021}.
\newblock \bibinfo{title}{{TransUNet}: Transformers make strong encoders for medical image segmentation}.
\newblock \bibinfo{journal}{arXiv preprint arXiv:2102.04306} .
\bibitem[{Chen et~al.(2018)Chen, Zhu, Papandreou, Schroff and Adam}]{chen2018encoder}
\bibinfo{author}{Chen, L.C.}, \bibinfo{author}{Zhu, Y.}, \bibinfo{author}{Papandreou, G.}, \bibinfo{author}{Schroff, F.}, \bibinfo{author}{Adam, H.}, \bibinfo{year}{2018}.
\newblock \bibinfo{title}{Encoder-decoder with atrous separable convolution for semantic image segmentation}, in: \bibinfo{booktitle}{ECCV}, pp. \bibinfo{pages}{801--818}.
\bibitem[{Chen et~al.(2023)Chen, Zhang, Wahle, Woo, Kassis, Kovarnik, Sonka and Lopez}]{chen2023deep}
\bibinfo{author}{Chen, Z.}, \bibinfo{author}{Zhang, H.}, \bibinfo{author}{Wahle, A.}, \bibinfo{author}{Woo, V.}, \bibinfo{author}{Kassis, N.}, \bibinfo{author}{Kovarnik, T.}, \bibinfo{author}{Sonka, M.}, \bibinfo{author}{Lopez, J.J.}, \bibinfo{year}{2023}.
\newblock \bibinfo{title}{Deep learning based automated optical coherence tomography analysis: A novel tool for identification of coronary artery lipid plaques and quantification of fibrous cap thickness}.
\newblock \bibinfo{journal}{Journal of the American College of Cardiology} \bibinfo{volume}{81}, \bibinfo{pages}{826--826}.
\bibitem[{Cui et~al.(2019)Cui, Jia, Lin, Song and Belongie}]{cui2019class}
\bibinfo{author}{Cui, Y.}, \bibinfo{author}{Jia, M.}, \bibinfo{author}{Lin, T.Y.}, \bibinfo{author}{Song, Y.}, \bibinfo{author}{Belongie, S.}, \bibinfo{year}{2019}.
\newblock \bibinfo{title}{Class-balanced loss based on effective number of samples}, in: \bibinfo{booktitle}{CVPR}, pp. \bibinfo{pages}{9268--9277}.
\bibitem[{Deng et~al.(2009)Deng, Dong, Socher, Li, Li and Fei-Fei}]{imagenet}
\bibinfo{author}{Deng, J.}, \bibinfo{author}{Dong, W.}, \bibinfo{author}{Socher, R.}, \bibinfo{author}{Li, L.J.}, \bibinfo{author}{Li, K.}, \bibinfo{author}{Fei-Fei, L.}, \bibinfo{year}{2009}.
\newblock \bibinfo{title}{{ImageNet}: A large-scale hierarchical image database}, in: \bibinfo{booktitle}{CVPR}, pp. \bibinfo{pages}{248--255}.
\bibitem[{Fan et~al.(2020)Fan, Ji, Zhou, Chen, Fu, Shen and Shao}]{fan2020pranet}
\bibinfo{author}{Fan, D.P.}, \bibinfo{author}{Ji, G.P.}, \bibinfo{author}{Zhou, T.}, \bibinfo{author}{Chen, G.}, \bibinfo{author}{Fu, H.}, \bibinfo{author}{Shen, J.}, \bibinfo{author}{Shao, L.}, \bibinfo{year}{2020}.
\newblock \bibinfo{title}{{PraNet}: Parallel reverse attention network for polyp segmentation}, in: \bibinfo{booktitle}{MICCAI, Part VI}, pp. \bibinfo{pages}{263--273}.
\bibitem[{Gessert et~al.(2018)Gessert, Lutz, Heyder, Latus, Leistner, Abdelwahed and Schlaefer}]{gessert2018automatic}
\bibinfo{author}{Gessert, N.}, \bibinfo{author}{Lutz, M.}, \bibinfo{author}{Heyder, M.}, \bibinfo{author}{Latus, S.}, \bibinfo{author}{Leistner, D.M.}, \bibinfo{author}{Abdelwahed, Y.S.}, \bibinfo{author}{Schlaefer, A.}, \bibinfo{year}{2018}.
\newblock \bibinfo{title}{Automatic plaque detection in {IVOCT} pullbacks using convolutional neural networks}.
\newblock \bibinfo{journal}{IEEE Transactions on Medical Imaging} \bibinfo{volume}{38}, \bibinfo{pages}{426--434}.
\bibitem[{He et~al.(2016)He, Zhang, Ren and Sun}]{resnet}
\bibinfo{author}{He, K.}, \bibinfo{author}{Zhang, X.}, \bibinfo{author}{Ren, S.}, \bibinfo{author}{Sun, J.}, \bibinfo{year}{2016}.
\newblock \bibinfo{title}{Deep residual learning for image recognition}, in: \bibinfo{booktitle}{CVPR}, pp. \bibinfo{pages}{770--778}.
\bibitem[{Jun et~al.(2019)Jun, Kang, Lee, Kweon, Na, Kang, Kim, Kim and Kim}]{Jun}
\bibinfo{author}{Jun, T.J.}, \bibinfo{author}{Kang, S.J.}, \bibinfo{author}{Lee, J.G.}, \bibinfo{author}{Kweon, J.}, \bibinfo{author}{Na, W.}, \bibinfo{author}{Kang, D.}, \bibinfo{author}{Kim, D.}, \bibinfo{author}{Kim, D.}, \bibinfo{author}{Kim, Y.H.}, \bibinfo{year}{2019}.
\newblock \bibinfo{title}{Automated detection of vulnerable plaque in intravascular ultrasound images}.
\newblock \bibinfo{journal}{Medical \& Biological Engineering \& Computing} \bibinfo{volume}{57}, \bibinfo{pages}{863--876}.
\bibitem[{Kolluru et~al.(2018)Kolluru, Prabhu, Gharaibeh, Bezerra, Guagliumi and Wilson}]{Kolluru}
\bibinfo{author}{Kolluru, C.}, \bibinfo{author}{Prabhu, D.}, \bibinfo{author}{Gharaibeh, Y.}, \bibinfo{author}{Bezerra, H.}, \bibinfo{author}{Guagliumi, G.}, \bibinfo{author}{Wilson, D.}, \bibinfo{year}{2018}.
\newblock \bibinfo{title}{Deep neural networks for {A}-line-based plaque classification in coronary intravascular optical coherence tomography images}.
\newblock \bibinfo{journal}{Journal of Medical Imaging} \bibinfo{volume}{5}, \bibinfo{pages}{044504--044504}.
\bibitem[{Kolodgie et~al.(2001)Kolodgie, Burke, Farb, Gold, Yuan, Narula, Finn and Virmani}]{kolodgie2001thin}
\bibinfo{author}{Kolodgie, F.D.}, \bibinfo{author}{Burke, A.P.}, \bibinfo{author}{Farb, A.}, \bibinfo{author}{Gold, H.K.}, \bibinfo{author}{Yuan, J.}, \bibinfo{author}{Narula, J.}, \bibinfo{author}{Finn, A.V.}, \bibinfo{author}{Virmani, R.}, \bibinfo{year}{2001}.
\newblock \bibinfo{title}{The thin-cap fibroatheroma: A type of vulnerable plaque: The major precursor lesion to acute coronary syndromes}.
\newblock \bibinfo{journal}{Current Opinion in Cardiology} \bibinfo{volume}{16}, \bibinfo{pages}{285--292}.
\bibitem[{Lee et~al.(2024)Lee, Kim, Dallan, Zimin, Hoori, Hassani, Makhlouf, Guagliumi, Bezerra and Wilson}]{lee2024deep}
\bibinfo{author}{Lee, J.}, \bibinfo{author}{Kim, J.N.}, \bibinfo{author}{Dallan, L.A.}, \bibinfo{author}{Zimin, V.N.}, \bibinfo{author}{Hoori, A.}, \bibinfo{author}{Hassani, N.S.}, \bibinfo{author}{Makhlouf, M.H.}, \bibinfo{author}{Guagliumi, G.}, \bibinfo{author}{Bezerra, H.G.}, \bibinfo{author}{Wilson, D.L.}, \bibinfo{year}{2024}.
\newblock \bibinfo{title}{Deep learning segmentation of fibrous cap in intravascular optical coherence tomography images}.
\newblock \bibinfo{journal}{Scientific Reports} \bibinfo{volume}{14}, \bibinfo{pages}{4393}.
\bibitem[{Lee et~al.(2022)Lee, Pereira, Gharaibeh, Kolluru, Zimin, Dallan, Kim, Hoori, Al-Kindi, Guagliumi et~al.}]{Lee}
\bibinfo{author}{Lee, J.}, \bibinfo{author}{Pereira, G.T.}, \bibinfo{author}{Gharaibeh, Y.}, \bibinfo{author}{Kolluru, C.}, \bibinfo{author}{Zimin, V.N.}, \bibinfo{author}{Dallan, L.A.}, \bibinfo{author}{Kim, J.N.}, \bibinfo{author}{Hoori, A.}, \bibinfo{author}{Al-Kindi, S.G.}, \bibinfo{author}{Guagliumi, G.}, et~al., \bibinfo{year}{2022}.
\newblock \bibinfo{title}{Automated analysis of fibrous cap in intravascular optical coherence tomography images of coronary arteries}.
\newblock \bibinfo{journal}{Scientific Reports} \bibinfo{volume}{12}, \bibinfo{pages}{21454}.
\bibitem[{Li and Jia(2019)}]{Li}
\bibinfo{author}{Li, L.}, \bibinfo{author}{Jia, T.}, \bibinfo{year}{2019}.
\newblock \bibinfo{title}{Optical coherence tomography vulnerable plaque segmentation based on deep residual {U-Net}}.
\newblock \bibinfo{journal}{Reviews in Cardiovascular Medicine} \bibinfo{volume}{20}, \bibinfo{pages}{171--177}.
\bibitem[{Lin et~al.(2017)Lin, Goyal, Girshick, He and Doll{\'a}r}]{lin2017focal}
\bibinfo{author}{Lin, T.Y.}, \bibinfo{author}{Goyal, P.}, \bibinfo{author}{Girshick, R.}, \bibinfo{author}{He, K.}, \bibinfo{author}{Doll{\'a}r, P.}, \bibinfo{year}{2017}.
\newblock \bibinfo{title}{Focal loss for dense object detection}, in: \bibinfo{booktitle}{ICCV}, pp. \bibinfo{pages}{2980--2988}.
\bibitem[{Liu et~al.(2019)Liu, Zhang, Zheng, Liu, Zhao and Yi}]{Liu}
\bibinfo{author}{Liu, R.}, \bibinfo{author}{Zhang, Y.}, \bibinfo{author}{Zheng, Y.}, \bibinfo{author}{Liu, Y.}, \bibinfo{author}{Zhao, Y.}, \bibinfo{author}{Yi, L.}, \bibinfo{year}{2019}.
\newblock \bibinfo{title}{Automated detection of vulnerable plaque for intravascular optical coherence tomography images}.
\newblock \bibinfo{journal}{Cardiovascular Engineering and Technology} \bibinfo{volume}{10}, \bibinfo{pages}{590--603}.
\bibitem[{Liu et~al.(2018)Liu, Deng, Xin, Zuo, Shi and Zheng}]{Liu_S}
\bibinfo{author}{Liu, S.}, \bibinfo{author}{Deng, Y.}, \bibinfo{author}{Xin, J.}, \bibinfo{author}{Zuo, W.}, \bibinfo{author}{Shi, P.}, \bibinfo{author}{Zheng, N.}, \bibinfo{year}{2018}.
\newblock \bibinfo{title}{{SRCNN}: Cardiovascular vulnerable plaque recognition with salient region proposal networks}, in: \bibinfo{booktitle}{2nd International Conference on Graphics and Signal Processing}, pp. \bibinfo{pages}{38--45}.
\bibitem[{Malakar et~al.(2019)Malakar, Choudhury, Halder, Paul, Uddin and Chakraborty}]{malakar2019review}
\bibinfo{author}{Malakar, A.K.}, \bibinfo{author}{Choudhury, D.}, \bibinfo{author}{Halder, B.}, \bibinfo{author}{Paul, P.}, \bibinfo{author}{Uddin, A.}, \bibinfo{author}{Chakraborty, S.}, \bibinfo{year}{2019}.
\newblock \bibinfo{title}{A review on coronary artery disease, its risk factors, and therapeutics}.
\newblock \bibinfo{journal}{Journal of Cellular Physiology} \bibinfo{volume}{234}, \bibinfo{pages}{16812--16823}.
\bibitem[{Min et~al.(2020)Min, Yoo, Kang, Lee, Cho, Lee, Ahn, Park, Lee, Kim et~al.}]{Min}
\bibinfo{author}{Min, H.S.}, \bibinfo{author}{Yoo, J.H.}, \bibinfo{author}{Kang, S.J.}, \bibinfo{author}{Lee, J.G.}, \bibinfo{author}{Cho, H.}, \bibinfo{author}{Lee, P.H.}, \bibinfo{author}{Ahn, J.M.}, \bibinfo{author}{Park, D.W.}, \bibinfo{author}{Lee, S.W.}, \bibinfo{author}{Kim, Y.H.}, et~al., \bibinfo{year}{2020}.
\newblock \bibinfo{title}{Detection of optical coherence tomography-defined thin-cap fibroatheroma in the coronary artery using deep learning}.
\newblock \bibinfo{journal}{EuroIntervention: Journal of EuroPCR} \bibinfo{volume}{16}, \bibinfo{pages}{404--412}.
\bibitem[{M{\"u}llner(2011)}]{mullner2011modern}
\bibinfo{author}{M{\"u}llner, D.}, \bibinfo{year}{2011}.
\newblock \bibinfo{title}{Modern hierarchical, agglomerative clustering algorithms}.
\newblock \bibinfo{journal}{arXiv preprint arXiv:1109.2378} .
\bibitem[{{National Center for Health Statistics}(2023)}]{national2023multiple}
\bibinfo{author}{{National Center for Health Statistics}}, \bibinfo{year}{2023}.
\newblock \bibinfo{title}{Multiple cause of death 2018--2021 on {CDC WONDER Database}}.
\newblock \bibinfo{journal}{Accessed February} \bibinfo{volume}{2}.
\bibitem[{Oktay et~al.(2018)Oktay, Schlemper, Folgoc, Lee, Heinrich, Misawa, Mori, McDonagh, Hammerla, Kainz et~al.}]{oktay2018attention}
\bibinfo{author}{Oktay, O.}, \bibinfo{author}{Schlemper, J.}, \bibinfo{author}{Folgoc, L.L.}, \bibinfo{author}{Lee, M.}, \bibinfo{author}{Heinrich, M.}, \bibinfo{author}{Misawa, K.}, \bibinfo{author}{Mori, K.}, \bibinfo{author}{McDonagh, S.}, \bibinfo{author}{Hammerla, N.Y.}, \bibinfo{author}{Kainz, B.}, et~al., \bibinfo{year}{2018}.
\newblock \bibinfo{title}{Attention {U-Net}: Learning where to look for the pancreas}.
\newblock \bibinfo{journal}{arXiv preprint arXiv:1804.03999} .
\bibitem[{Paszke et~al.(2019)Paszke, Gross, Massa, Lerer, Bradbury, Chanan, Killeen, Lin, Gimelshein, Antiga et~al.}]{paszke2019pytorch}
\bibinfo{author}{Paszke, A.}, \bibinfo{author}{Gross, S.}, \bibinfo{author}{Massa, F.}, \bibinfo{author}{Lerer, A.}, \bibinfo{author}{Bradbury, J.}, \bibinfo{author}{Chanan, G.}, \bibinfo{author}{Killeen, T.}, \bibinfo{author}{Lin, Z.}, \bibinfo{author}{Gimelshein, N.}, \bibinfo{author}{Antiga, L.}, et~al., \bibinfo{year}{2019}.
\newblock \bibinfo{title}{{PyTorch}: An imperative style, high-performance deep learning library}.
\newblock \bibinfo{journal}{Advances in Neural Information Processing Systems} \bibinfo{volume}{32}, \bibinfo{pages}{8024--8035}.
\bibitem[{Ronneberger et~al.(2015)Ronneberger, Fischer and Brox}]{ronneberger2015u}
\bibinfo{author}{Ronneberger, O.}, \bibinfo{author}{Fischer, P.}, \bibinfo{author}{Brox, T.}, \bibinfo{year}{2015}.
\newblock \bibinfo{title}{{U-Net}: Convolutional networks for biomedical image segmentation}, in: \bibinfo{booktitle}{MICCAI, Part III}, pp. \bibinfo{pages}{234--241}.
\bibitem[{Shi et~al.(2023)Shi, Xin, Du, Wu, Deng, Cai and Zheng}]{shi2023automatic}
\bibinfo{author}{Shi, P.}, \bibinfo{author}{Xin, J.}, \bibinfo{author}{Du, S.}, \bibinfo{author}{Wu, J.}, \bibinfo{author}{Deng, Y.}, \bibinfo{author}{Cai, Z.}, \bibinfo{author}{Zheng, N.}, \bibinfo{year}{2023}.
\newblock \bibinfo{title}{Automatic lumen and anatomical layers segmentation in {IVOCT} images using meta learning}.
\newblock \bibinfo{journal}{Journal of Biophotonics} \bibinfo{volume}{16}, \bibinfo{pages}{e202300059}.
\bibitem[{Shi et~al.(2018)Shi, Xin, Liu, Deng and Zheng}]{shi2018vulnerable}
\bibinfo{author}{Shi, P.}, \bibinfo{author}{Xin, J.}, \bibinfo{author}{Liu, S.}, \bibinfo{author}{Deng, Y.}, \bibinfo{author}{Zheng, N.}, \bibinfo{year}{2018}.
\newblock \bibinfo{title}{Vulnerable plaque recognition based on attention model with deep convolutional neural network}, in: \bibinfo{booktitle}{40th Annual International Conference of the IEEE Engineering in Medicine and Biology Society (EMBC)}, pp. \bibinfo{pages}{834--837}.
\bibitem[{Tsao et~al.(2022)Tsao, Aday, Almarzooq, Alonso, Beaton, Bittencourt, Boehme, Buxton, Carson, Commodore-Mensah et~al.}]{tsao2022heart}
\bibinfo{author}{Tsao, C.W.}, \bibinfo{author}{Aday, A.W.}, \bibinfo{author}{Almarzooq, Z.I.}, \bibinfo{author}{Alonso, A.}, \bibinfo{author}{Beaton, A.Z.}, \bibinfo{author}{Bittencourt, M.S.}, \bibinfo{author}{Boehme, A.K.}, \bibinfo{author}{Buxton, A.E.}, \bibinfo{author}{Carson, A.P.}, \bibinfo{author}{Commodore-Mensah, Y.}, et~al., \bibinfo{year}{2022}.
\newblock \bibinfo{title}{Heart disease and stroke statistics—2022 update: A report from the {American Heart Association}}.
\newblock \bibinfo{journal}{Circulation} \bibinfo{volume}{145}, \bibinfo{pages}{e153--e639}.
\bibitem[{Vaswani et~al.(2017)Vaswani, Shazeer, Parmar, Uszkoreit, Jones, Gomez, Kaiser and Polosukhin}]{vaswani2017attention}
\bibinfo{author}{Vaswani, A.}, \bibinfo{author}{Shazeer, N.}, \bibinfo{author}{Parmar, N.}, \bibinfo{author}{Uszkoreit, J.}, \bibinfo{author}{Jones, L.}, \bibinfo{author}{Gomez, A.N.}, \bibinfo{author}{Kaiser, {\L}.}, \bibinfo{author}{Polosukhin, I.}, \bibinfo{year}{2017}.
\newblock \bibinfo{title}{Attention is all you need}.
\newblock \bibinfo{journal}{Advances in Neural Information Processing Systems} \bibinfo{volume}{30}, \bibinfo{pages}{6000–6010}.
\bibitem[{Wang et~al.(2012)Wang, Chamie, Bezerra, Yamamoto, Kanovsky, Wilson, Costa and Rollins}]{wang2012volumetric}
\bibinfo{author}{Wang, Z.}, \bibinfo{author}{Chamie, D.}, \bibinfo{author}{Bezerra, H.G.}, \bibinfo{author}{Yamamoto, H.}, \bibinfo{author}{Kanovsky, J.}, \bibinfo{author}{Wilson, D.L.}, \bibinfo{author}{Costa, M.A.}, \bibinfo{author}{Rollins, A.M.}, \bibinfo{year}{2012}.
\newblock \bibinfo{title}{Volumetric quantification of fibrous caps using intravascular optical coherence tomography}.
\newblock \bibinfo{journal}{Biomedical Optics Express} \bibinfo{volume}{3}, \bibinfo{pages}{1413--1426}.
\bibitem[{Wang et~al.(2023)Wang, Shao, Sun, Huang, Wang, Li, Li and Yu}]{wang2023vision}
\bibinfo{author}{Wang, Z.}, \bibinfo{author}{Shao, Y.}, \bibinfo{author}{Sun, J.}, \bibinfo{author}{Huang, Z.}, \bibinfo{author}{Wang, S.}, \bibinfo{author}{Li, Q.}, \bibinfo{author}{Li, J.}, \bibinfo{author}{Yu, Q.}, \bibinfo{year}{2023}.
\newblock \bibinfo{title}{Vision {Transformer} based multi-class lesion detection in {IVOCT}}, in: \bibinfo{booktitle}{International Conference on Medical Image Computing and Computer-Assisted Intervention}, \bibinfo{organization}{Springer}. pp. \bibinfo{pages}{327--336}.
\bibitem[{Zahnd et~al.(2017)Zahnd, Hoogendoorn, Combaret, Karanasos, P{\'e}ry, Sarry, Motreff, Niessen, Regar, Van~Soest et~al.}]{zahnd2017contour}
\bibinfo{author}{Zahnd, G.}, \bibinfo{author}{Hoogendoorn, A.}, \bibinfo{author}{Combaret, N.}, \bibinfo{author}{Karanasos, A.}, \bibinfo{author}{P{\'e}ry, E.}, \bibinfo{author}{Sarry, L.}, \bibinfo{author}{Motreff, P.}, \bibinfo{author}{Niessen, W.}, \bibinfo{author}{Regar, E.}, \bibinfo{author}{Van~Soest, G.}, et~al., \bibinfo{year}{2017}.
\newblock \bibinfo{title}{Contour segmentation of the intima, media, and adventitia layers in intracoronary {OCT} images: Application to fully automatic detection of healthy wall regions}.
\newblock \bibinfo{journal}{International Journal of Computer Assisted Radiology and Surgery} \bibinfo{volume}{12}, \bibinfo{pages}{1923--1936}.
\bibitem[{Zahnd et~al.(2015)Zahnd, Karanasos, Van~Soest, Regar, Niessen, Gijsen and van Walsum}]{Zahnd}
\bibinfo{author}{Zahnd, G.}, \bibinfo{author}{Karanasos, A.}, \bibinfo{author}{Van~Soest, G.}, \bibinfo{author}{Regar, E.}, \bibinfo{author}{Niessen, W.}, \bibinfo{author}{Gijsen, F.}, \bibinfo{author}{van Walsum, T.}, \bibinfo{year}{2015}.
\newblock \bibinfo{title}{Quantification of fibrous cap thickness in intracoronary optical coherence tomography with a contour segmentation method based on dynamic programming}.
\newblock \bibinfo{journal}{International Journal of Computer Assisted Radiology and Surgery} \bibinfo{volume}{10}, \bibinfo{pages}{1383--1394}.

\end{thebibliography}

\end{document}